\documentclass[%
 reprint,
 superscriptaddress,
 amsmath,amssymb,
 aps,
floatfix,
]{revtex4-2}

\usepackage{graphicx}
\usepackage{dcolumn}
\usepackage{bm}

\begin{document}

\preprint{APS/123-QED}

 \title{Systematic Modification of Functionality in Disordered Elastic Networks Through Free Energy Surface Tailoring}

\author{Dan Mendels}
 \affiliation{%
 Pritzker School of Molecular Engineering, University of Chicago, 5640 S. Ellis Avenue, Chicago, Illinois USA 60637
}
\author{Fabian Byl\' ehn}%
\affiliation{%
 Pritzker School of Molecular Engineering, University of Chicago, 5640 S. Ellis Avenue, Chicago, Illinois USA 60637
}
\author{Timothy W. Sirk}
\affiliation{%
 Polymers Branch, U.S. CCDC Army Research Laboratory, Aberdeen Proving Ground, Maryland 21005, United States
}
\author{Juan J. de Pablo}
\email{depablo@uchicago.edu} 
\affiliation{%
 Pritzker School of Molecular Engineering, University of Chicago, 5640 S. Ellis Avenue, Chicago, Illinois USA 60637
}%




\begin{abstract}
Advances in manufacturing and characterization of complex molecular systems have created a need for new methods for design at molecular length scales. Emerging approaches are increasingly relying on the use of Artificial Intelligence (AI), and the training of AI models on large data libraries. This paradigm shift has led to successful applications, but shortcomings related to interpretability and generalizability continue to pose challenges. Here, we explore an alternative paradigm in which AI is combined with physics-based considerations for molecular and materials engineering. Specifically, collective variables, akin to those used in enhanced sampled simulations, are constructed using a machine learning (ML) model trained on data gathered from a single system. Through the ML-constructed collective variables, it becomes possible to identify critical molecular interactions in the system of interest, the modulation of which enables a systematic tailoring of the system's free energy landscape. To explore the efficacy of the proposed approach, we use it to engineer allosteric regulation, and uniaxial strain fluctuations in a complex disordered elastic network. Its successful application in these two cases provides insights regarding how functionality is governed in systems characterized by extensive connectivity, and points to its potential for design of complex molecular systems.  
\end{abstract}

\maketitle

\newpage
\section*{Introduction}

Progress in the manufacturing and characterization of complex molecular and material systems is often hampered by the complexity of such systems and the enormity of the available design space. Engineering systems at molecular length scales remains a challenging, costly, and time-consuming endeavour. There is a need for new design and optimization methods that can, on the one hand, harness the underlying complexity, and, on the other, identify the regions in design space that are most likely to provide fruitful solutions for a given problem. 

Several promising strategies for devising such methods revolve around the use of artificial intelligence (AI) and particularly its application to large libraries of data associated with sets of diverse systems corresponding to design problems being considered. Given the increasing amounts of data being generated through experiments and simulations, the use of AI for molecular and materials design in this way continues to grow. There are, however, limitations to such approaches' effectiveness. These include (i) the volume of training data that is required, (ii) the inability to interpret certain outcomes given the black-box nature of many AI-based algorithms, and (iii) the limited applicability of the constructed models beyond the underlying training domain. While overcoming these issues is a matter of on-going research within the field of AI at large, we have recently proposed a machine learning-based method designed to circumvent them. The method, referred to as Collective Variables for Free Energy Surface Tailoring (CV-FEST), does so by (a) using the considerable amount of data generated in simulations or experiments of a single system, and by (b) relying on the powerful ability of certain machine learning algorithms to generate insightful dimensionally-reduced representations of complex high-dimensional data.

Specifically, CV-FEST relies on the notion that the functionality of many systems can often be characterized using a dimensionally reduced representation of their free energy surfaces (FES) within a space spanned by a set of collective variables (CVs). Such a depiction serves two interconnected objectives: (i) it provides insight into the mechanisms underlying the functionality of a considered system, and (ii) it allows to condense the most essential information about such a system into a small number of parameters that can be tuned with the aid of optimization algorithms (e.g. \cite{long2018rational,white2014efficient, amirkulova2019recent, gil2016empirical, marinelli2015ensemble, cesari2018using}) for design purposes. 

In ref. \cite{mendels2022collective} we focused on analysing and modifying the functionality of systems that consist of relatively small numbers of degrees of freedom, e.g. a small peptide. Here, we focus on a system of much higher complexity and focus on the question of whether its underlying FES can be manipulated at will. Specifically, we consider an elastic network consisting of roughly 1000 harmonic bonds. Elastic networks have become a subject of increasing interest in recent years given their functional similarities to proteins \cite{bahar2010global, rocks2017designing} and their ability to exhibit metamaterial qualities \cite{reid2018auxetic, hexner2020periodic}. In addition, elastic networks form a natural framework for studying network structure and behavior in the physical realm, potentially providing new and general perspectives on network behavior \cite{albert2002statistical, liu2016control}. 

\begin{figure}
	\centering
    \includegraphics[trim=0 293 76 313,clip, width=0.96\columnwidth]{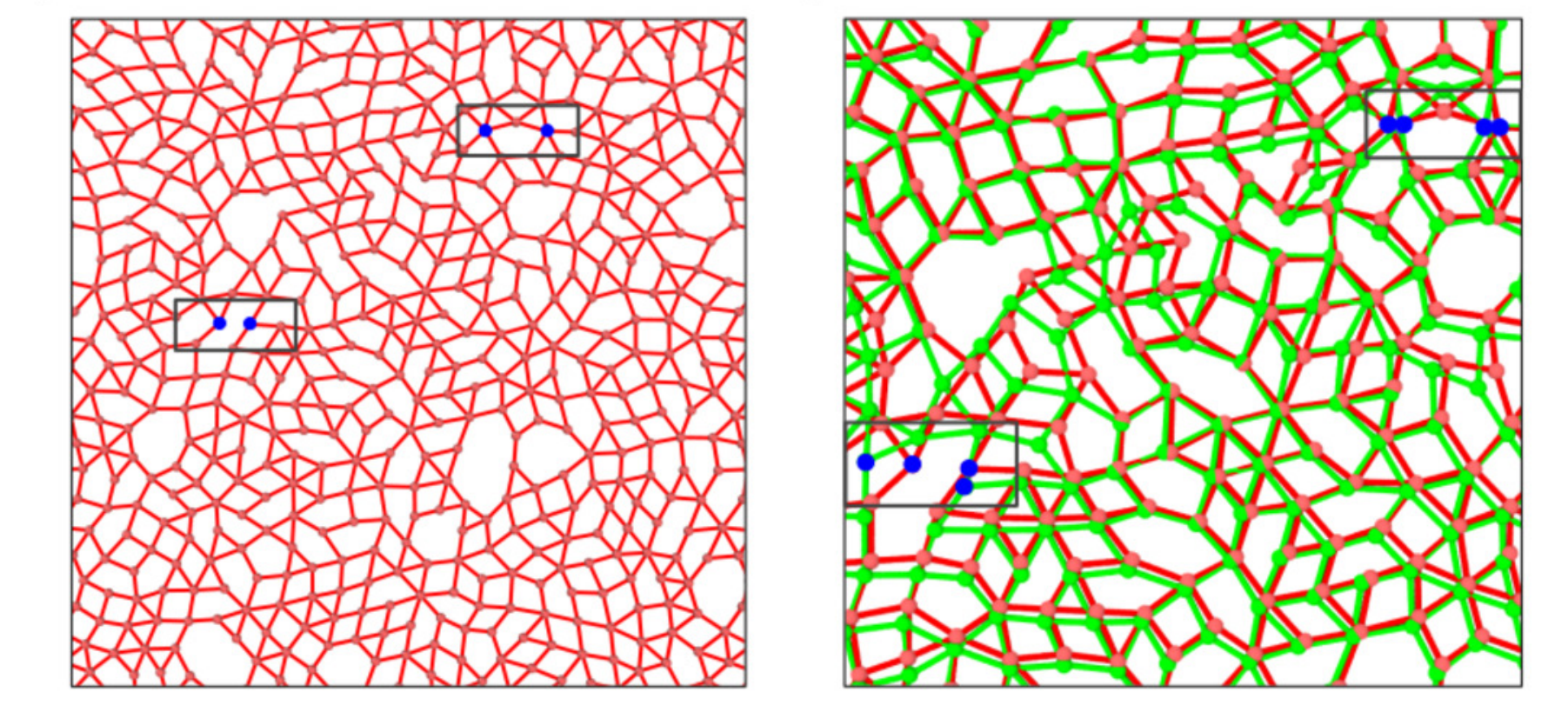}   
    \caption{(a) Illustration of a simulated network from which 12 bonds were removed to embed allosteric response between the indicated sites. The source nodes are represented by the blue circles while the target nodes represented by the green circles. (b) Magnified view of the network in its activated state (in green) overlayed with the network in its inactivated state (in red). The source beads are pinched towards one another in the activated state leading to the distancing of the target beads from one another. 
    }
    \label{fig:Network illustration}
\end{figure}

For concreteness, we consider a two-dimensional disordered network consisting of identical beads connected by harmonic bonds of identical elastic modulus (see Fig. \ref{fig:Network illustration} for illustration). We focus on two properties of the network. The first, allosteric regulation, is analogous to that found in proteins, in which the conformation of a target site (referred to as the active site) in the network is regulated by the conformation of a different, distant site in the network (referred to as the allosteric site). The second functionality examined here is the network's uniaxial mechanical behavior, as manifested by its uniaxial strain fluctuations. We find that using CV-FEST, we are able to modify these two network characteristics in a simple and tractable manner by systemically tailoring the FES associated with them.

\subsubsection*{Collective Variables for Free Energy Tailoring}

As described above, CV-FEST relies on the idea that the functionality of many systems in nature can be captured using a dimensionally reduced representation of their free energy surface (FES) within a space spanned by a set of collective variables (CVs). The CVs describe the relevant modes of behavior of a system, similar to the role they play in enhanced sampling methods such as umbrella sampling \cite{Torrie1977}, metadynamics \cite{laio_parrinello_2002} and adaptive biasing force \cite{darve2001calculating}, which are used to accelerate the sampling of systems for which the characteristic time scales lie beyond the reach of ordinary simulations. Generally speaking, CVs are functions of the system's atomic coordinates $\mathbf{s}(\mathbf{R})$ and can be defined through the following relations: 

\begin{equation}
\label{collective variables defined}
P(\mathbf{s})=\int{d\mathbf{R}\delta[\mathbf{s}-\mathbf{s(\mathbf{R})}}] P(\mathbf{R})
\end{equation}

\noindent where $P(\mathbf{s})$ is the system probability to hold a set  of CV values $\mathbf{s}$, $P(\mathbf{R})$ is the Boltzmann probability, and $\delta$ is Dirac's delta function.
A system's FES with respect to the utilized set of CVs then follows:

\begin{equation}
\label{FES for CVs}
F(\mathbf{s})=-\frac{1}{\beta} \log P(\mathbf{s})
\end{equation}

\noindent where $\beta=1\slash k_BT$ and $k_B$ is Boltzmann's constant and $T$ is the temperature.

While the construction of adequate CVs in the context of enhanced sampling can be challenging, once achieved, CVs encode the physical essence of the processes that determine a system's behavior at long time scales. This renders them potentially useful tools for engineering. While traditionally the construction of CVs was deemed to require a certain degree of expertise regarding the system being studied, new machine learning based methods for this task have been introduced in recent years \cite{ribeiro2018reweighted,mendels2018collective,bonati2020data,chen2018collective,wehmeyer2018time,sultan2018automated,mccarty2017variational}. CV-FEST utilizes Harmonic Linear Discriminant Analysis (HLDA) \cite{mendels2018collective,piccini2018metadynamics,mendels2018folding,rizzi2019blind,zhang2019improving}, given its ease of use and straightforward interpretation. HLDA constructs CVs as linear weighted sums of descriptors and requires as input a limited amount of information, which can be collected via short simulations in states relevant to the processes being considered. Given their linear form, the interpretation of the HLDA CVs is straightforward; descriptors attaining larger weights in absolute value are deemed to be associated with the forces that encompass higher physicochemical importance with respect to the relevant behavior or process. The HLDA CVs descriptor hierarchy can thus be used to identify the set of forces and interactions in a system that can be tuned for the purposes of tailoring its FES, and modifying its functionality in desirable ways.

While HLDA was originally designed to construct CVs that correspond to rare transitions occurring between metastable states \cite{mendels2018collective,piccini2018metadynamics}, its applicability has since been shown to extend also to cases in which the input data is collected in unstable states \cite{brotzakis2019augmented}. In what follows, we take advantage of this aspect of the method. The application of HLDA requires that a list of system descriptors $d_i$ be identified as an input, e.g. distances between beads, bond angles, or more complex variables such as the enthalpy or entropy of a system  \cite{piaggi2017enhancing,Mendels2018}. In the current context, however, we limit the type of descriptors to those which directly correspond to tunable force potentials of the system, namely the system's bond potentials.

Once a descriptor set is assembled, HLDA requires as input the expectation value vectors $\mu_I$ and covariance matrices $\Sigma_I$ with respect to the predefined descriptor space, corresponding to each of the states $I\in{M}$ associated with the relevant processes. The computation of the elements can be carried out using data collected in short unbiased simulations of each of the relevant states. To construct the CVs, HLDA estimates the directions $\mathbf{W}$ in the $N_d$ dimensional descriptor space on which the projections of the collected training distributions are best separated. This is done through the maximization of the ratio between the training data, the so-called between-class $\mathbf{S}_b$ and within-class $\mathbf{S}_w$ scatter matrices, and can be written as:

\begin{equation}
\label{fisher_ration}
\mathcal{J(\mathbf{W})} = \frac{\mathbf{W}^T \mathbf{S}_b \mathbf{W}}{\mathbf{W}^T \mathbf{S}_w \mathbf{W}}
\end{equation}

\noindent
with 

\begin{equation}
\label{between_class}
\mathbf{S}_b = \left( \boldsymbol{\mu}_I - \boldsymbol{\bar{\mu}} \right)\left( \boldsymbol{\bar{\mu}} - \boldsymbol{\mu}_I \right)^T
\end{equation}

\noindent
where $\boldsymbol{\mu}_I$ are the expectation value vectors of the I-th metastable state and $\boldsymbol{\bar{\mu}}$ is the overall mean of the distributions, i.e. $\boldsymbol{\bar{\mu}}=1/M \sum_{I=1}^{M}$, and with 

\begin{equation}
\label{harmonic_mean}
\mathbf{S}_w = \frac{1}{\frac{1}{\boldsymbol{\Sigma}_1} + \frac{1}{\boldsymbol{\Sigma}_2}+...+\frac{1}{\boldsymbol{\Sigma}_M}}.
\end{equation}. 

\noindent
Given the normalization $\mathbf{W}^T \mathbf{S}_w \mathbf{W}=1$, Eq. \ref{fisher_ration} can be shown to be equivalent to solving the eigenvalue equation \cite{piccini2018metadynamics}:

\begin{equation}
\label{maximizer}
\mathbf{S}_w^{-1}\mathbf{S}_b \mathbf{W}=\lambda\mathbf{W} 
\end{equation}

\noindent
The eigenvectors of Eq. \ref{maximizer} associated with the largest M-1 eigenvalues define the directions in the $N_d$ space along which the distributions obtained from the M sampled states overlap the least, and thus constitute the CVs that correspond to the transitions of interest. Using these CVs, we can now systematically tailor the system FES and modify its functionality in a purposeful manner. To do that, the leading descriptors of each of the constructed CVs are identified and their corresponding interaction potentials are changed.

\subsection*{Results}
\subsubsection*{Allosteric response}

The first network functionality we focus on is that of allosteric regulation. The term allosteric regulation originates from the study of proteins, referring to the alteration of the activity of a site in the protein (e.g., its ability to have a ligand bind to it) through the binding of an effector molecule to another distal site in it, i.e. the allosteric site. Allosteric regulation plays an important role in many biological processes, such as transcriptional regulation and metabolism, and is rooted in the fundamental physical properties of macromolecular systems. Interestingly, its underlying mechanisms are still poorly understood and hence it constitutes a central theme of present research in the field of biology \cite{wodak2019allostery, faure2022mapping}.  

In the considered network, we simulate the binding of an effector molecule to an allosteric site as the pinching of two neighboring non-bonded beads towards each other. We refer to these two beads as the source beads (represented by blue circles in Fig. \ref{fig:Network illustration}a) of the considered allosteric process. Correspondingly, a pair of target beads is defined at a different location in the network, representing the active site (also represented by blue circles in Fig. \ref{fig:Network illustration}a). The allosteric activation of the target beads, defined as their distancing from one another in response to the pinching of the source beads, is embedded into the network using the "tuning by pruning" algorithm introduced in ref. \cite{rocks2017designing} (see Methods for more details). Thus, when the source beads are at their relaxed positions, the distance between the target beads assumes its initial value and the active site is in its inactivated state. However, when the source beads are pinched beyond a threshold, the target beads respond by distancing from one another, and the active site is deemed to be activated. We consider an additional third state of the system, corresponding to the phenomenon of negative cooperativity \cite{hunter2009cooperativity}. A scenario is considered in which the binding of a ligand to a site neighboring the allosteric site inhibits the binding of the effector molecule, for example by steric repulsion. We simulate this scenario as a state of the network in which the source beads are stretched away from one another, thus inhibiting their ability to arrive to the pinched active state.  The system in practice can thus be in one of three states: the inactivated, activated, or inhibited state.

To explore our ability to systematically tailor the system's FES associated with its allosteric behavior, we start by defining a descriptor set for the problem at hand. In this case we opt for a descriptor set composed of all the network bond distances. Next, we run short simulations in the three different states of the system at finite temperature, constraining the source beads to their respective positions at each state. Using the data collected in the simulations we then compute the expectation values $\mu_I$ and covariance matrices $\Sigma_I$  associated with each of the states, and using Eq. \ref{maximizer} we compute the HLDA CVs corresponding to the three-state system. The weight distribution corresponding to the first eigenvector is presented in the inset of Fig. \ref{fig:allosteric response results}c.

\begin{figure}
	\centering
    \includegraphics[trim=0 0 140 0,clip, width=0.96\columnwidth]{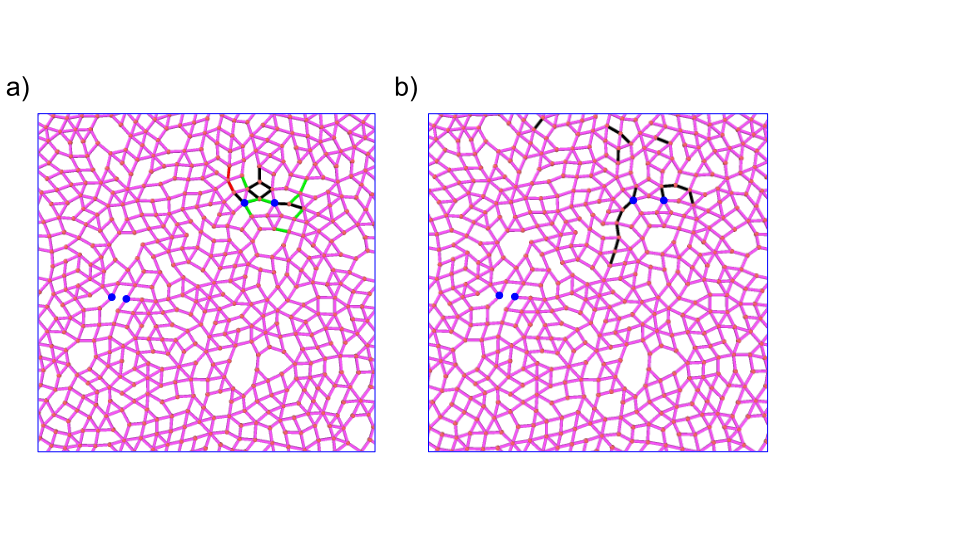}   
    \caption{(a) Depiction of the 10 highest weighted bonds in absolute value of the calculated HLDA CVs. The top bonds corresponding to  HLDA\textsubscript{1} are shown in black, and the top bonds corresponding to  HLDA\textsubscript{2} are shown in green. The top bonds corresponding to both HLDA\textsubscript{1} and  HLDA\textsubscript{2} are red. (b) Depiction of the bonds ranked 12-27 in the weight hierarchy of  HLDA\textsubscript{1}. Decreasing the bond coefficient of these bonds inhibits the activation of the active site.
    }
    \label{fig:Selected_bonds of allosteric response illustrated }
\end{figure}

\begin{figure}
	\centering
    \includegraphics[trim=0 0 190 0,clip, width=0.96\columnwidth]{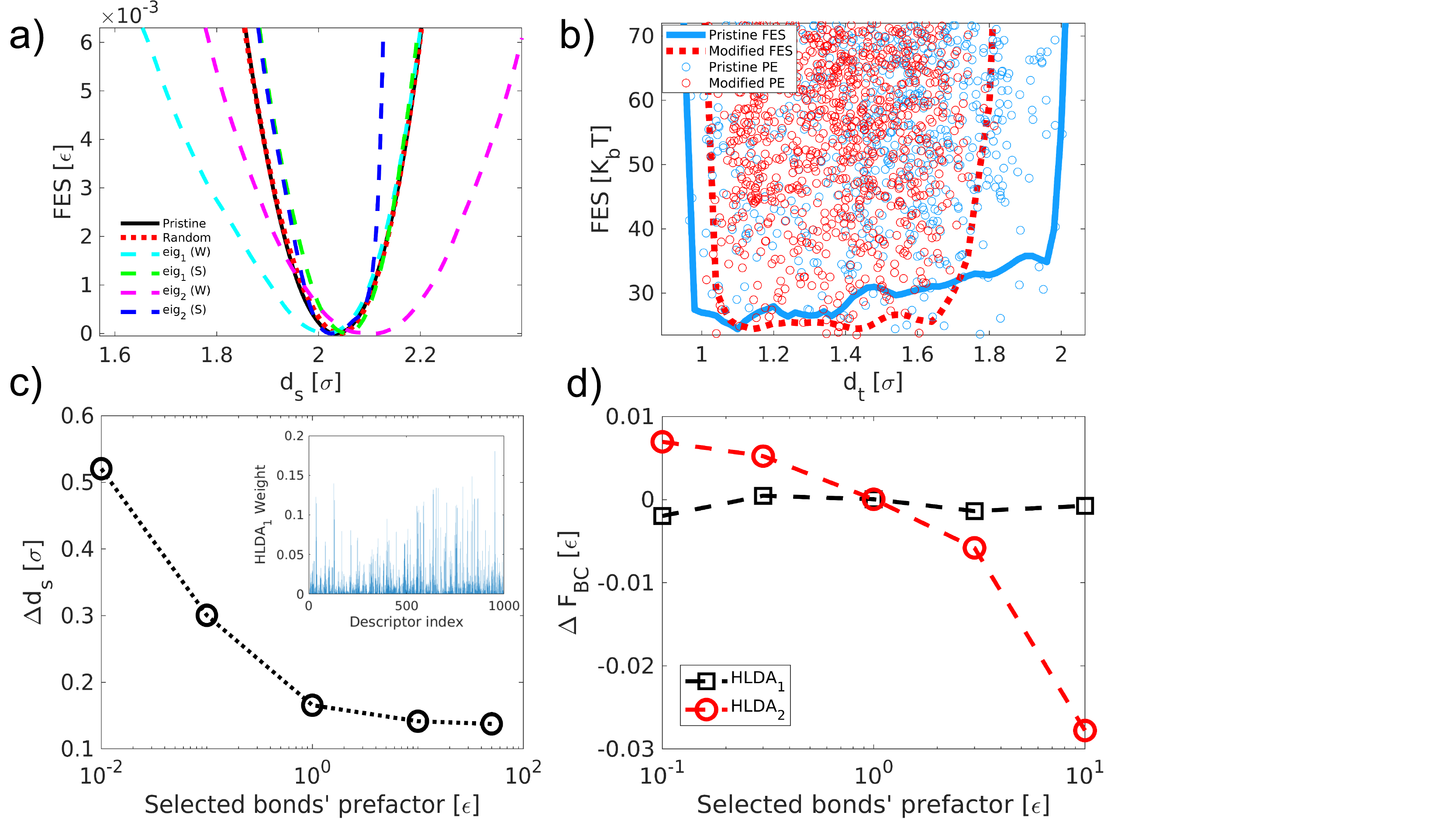}   
    \caption{(a) FES with different realizations of the allosteric network as a function of the distance between the source nodes. FES of the pristine network (solid line), and FES obtained after 30 random bonds are strengthened to $50 \epsilon$ (doted line). Also shown: FES obtained of networks in which the 10 highest weighted bonds of either HLDA\textsubscript{1} or HLDA\textsubscript{2} are either strengthened (S) to $50 \epsilon$ or weakened (W) to $0.1 \epsilon$ (dashed lines). (b) Potential energy distributions and reweighted FES as a function of $d_t$ calculated using the methodology of \cite{tiwary2015time}. The pristine network can reach the fully activated state at $d_t>1.7 \sigma$, while this state is not accessible to the network in which the bonds ranked 12-27 in the HLDA\textsubscript{1} weight hierarchy are weakened. (c) Minimal pinching length for which activation of the target beads is initiated as a function of the bond coefficient of the 10 highest weighted HLDA\textsubscript{1}  bonds. Inset: Absolute value of the weight distribution of HLDA\textsubscript{1}, generated in the first iteration of the calculation. (d) $\Delta F_{BC}$ (calculated using Eq. S10), the free energy difference between state C (the stretched state), $d_s=2.19 \sigma$, and the pinched state (defined as the point at which activation of the target site is initiated) as a function of the bond coefficient of the targeted bonds.  
    }
    \label{fig:allosteric response results}
\end{figure}

While the HLDA eigenvectors corresponding to the top two eigenvalues provide good separation between the predefined states, to augment our ability to tailor the system's FES, prior to selecting the top weighted descriptors we apply a two-dimensional rotation to the plane spanned by the two HLDA CVs. This is implemented in such a way that states A and B are best separated with respect to the direction corresponding to the rotated HLDA\textsubscript{1}, while states B and C are best separated with respect to the direction corresponding to the rotated HLDA\textsubscript{2} (see Fig. S5 in the SI for illustration). In analogy to the concept of normal modes, applying the described rotation allows us to substantially decouple the effects induced by the  modification of bonds associated with HLDA\textsubscript{1} on the FES associated with those induced by the modification of the bonds involved in HLDA\textsubscript{2}.  In essence, modifying bonds associated with HLDA\textsubscript{1} predominantly affect the free energy difference between states A and B, whereas modifying bonds associated with HLDA\textsubscript{2} predominantly affects the free energy difference between states A and C, as can be seen in Fig. \ref{fig:allosteric response results}a. Given the linearity of HLDA, the resulting CVs tend to be dominated by the descriptors associated with the largest weights in the CV, rendering the weight distribution of the lower weighted descriptors potentially less accurate. To circumvent this issue, we apply HLDA iteratively, whereby after each iteration (and rotation), the top five weighted descriptors (in absolute value) of each of the two constructed CVs are selected and removed from the descriptor list for subsequent iterations. The top bonds selected in this way are highlighted in Fig. \ref{fig:Selected_bonds of allosteric response illustrated }a. As can be seen, both bond sets corresponding to the two HLDA CVs form distinct patterns, reflecting their functional significance.

To test our ability to tailor the FES corresponding to the allosteric and cooperative behaviors of the system, we alter the bond coefficient of the selected bonds and then compute the FES.  Given the large differences in free energy between the considered states, the computation of the system's FES requires the use of an enhanced sampling approach. Here we use Well Tempered Metadynamics (WTMD) \cite{Barducci2008}, using the distance between the source beads $d_s$ as the biasing CV for simulations. (See the Methods section for details and Fig. S6 for an example of the time-dependent behavior of the source and target nodes in such simulations). Upon convergence of the WTMD runs, we compute the FES of the system using Eq. \ref{Eq: WTMD}. Fig. \ref{fig:allosteric response results}a presents the FES computed in this way for several realizations of the system. One can appreciate that altering the bond coefficient of the selected bonds gives rise to significant changes in the system's FES. In contrast, similar alterations of randomly selected bonds don't lead to noticeable changes of the FES. It can also be seen that altering the bonds associated with HLDA\textsubscript{1} mainly affects the FES branch corresponding to the $A \leftrightarrow B$ transition, while altering those associated with HLDA\textsubscript{2}, predominantly affects the FES branch corresponding to the $B \leftrightarrow C$ transition. 

An examination of the effects of altering the bond coefficient of the top ten bonds corresponding to HLDA\textsubscript{1} reveals that one can systematically modify the extent to which the source nodes need to be pinched in order to initiate the ''activation'' of the target beads. Fig. \ref{fig:allosteric response results}c illustrates this feature by showing the dependence of the distance $d_s$ for which the allosteric response is initiated as a function of the bond coefficient of the top ten HLDA\textsubscript{1} bonds. Our results show that as the selected bonds are weakened, the source beads need to be brought closer to one another for allosteric activation to occur, and vice versa. One can envision inducing similar effects in real proteins; by systemically softening the environment of an allosteric site, one could alter the types of molecules (e.g., different dimensions or interactions with the allosteric site) that would lead to its activation. 

Interestingly, while altering the top bonds associated with HLDA\textsubscript{1} has a significant effect on the distance $d_S$ for which the allosteric response is initiated, we find that it induces a more modest effect on the free energy difference between states A and B, and consequently also on that between states B and C. In contrast, however, modifying the bonds associated with HLDA\textsubscript{2} leads to large changes in the free energy difference between states B and C, thereby allowing us to easily alter this free energy difference between the activated and inhibited states, as shown in Fig. \ref{fig:allosteric response results}d. 

By repeating the bond selection procedure for several iterations we find that we can reveal segments that compose the primary channel of mechanical communication between the allosteric and active sites. The corresponding bonds, placed in positions 12-27 in the hierarchy of HLDA\textsubscript{1}, are highlighted in Fig. \ref{fig:Selected_bonds of allosteric response illustrated }b. Specifically, we find that weakening this set of bonds limits the communication between the sites, precluding the activation of the active site, as illustrated in Fig. S7. Fig. \ref{fig:allosteric response results}b illustrates this point further by presenting the FES corresponding to the distance between the target beads, plotted for the pristine system and the modified system. It can be seen that the fully activated region corresponding to $d_t>1.7\sigma$ is inaccessible in the case of the modified system. Considering again the analogy to proteins, it would be intriguing to explore if the proposed methodology would be able to help shed light on the prominent question of how communication occurs between allosteric and active sites.

\subsubsection*{Uniaxial Strain Fluctuations}

To complement the analysis of the network's allosteric behavior, we apply CV-FEST also to a comparably more global attribute of the network, namely its uniaxial strain fluctuations \cite{parrinello1982strain}. The network is constructed in the same manner as before. As in the previous case, in order to systematically modify the behavior of the network, we start by collecting training data. We do this by running a slow uniaxial compression simulation at constant temperature and constant lateral dimension. The network is deformed in the x direction to a strain of $\delta L_x/L_x=0.0075$. To construct the HLDA CV that corresponds to the network's strain fluctuations in the x direction, we use data collected in two short segments of the compression simulation. The first, taken from the beginning of the simulation when the network is nearly relaxed, and the second from the end of the simulation when the network is nearly fully deformed. 

\begin{figure}
	\centering
    \includegraphics[trim=0 100 0 0,clip, width=0.96\columnwidth]{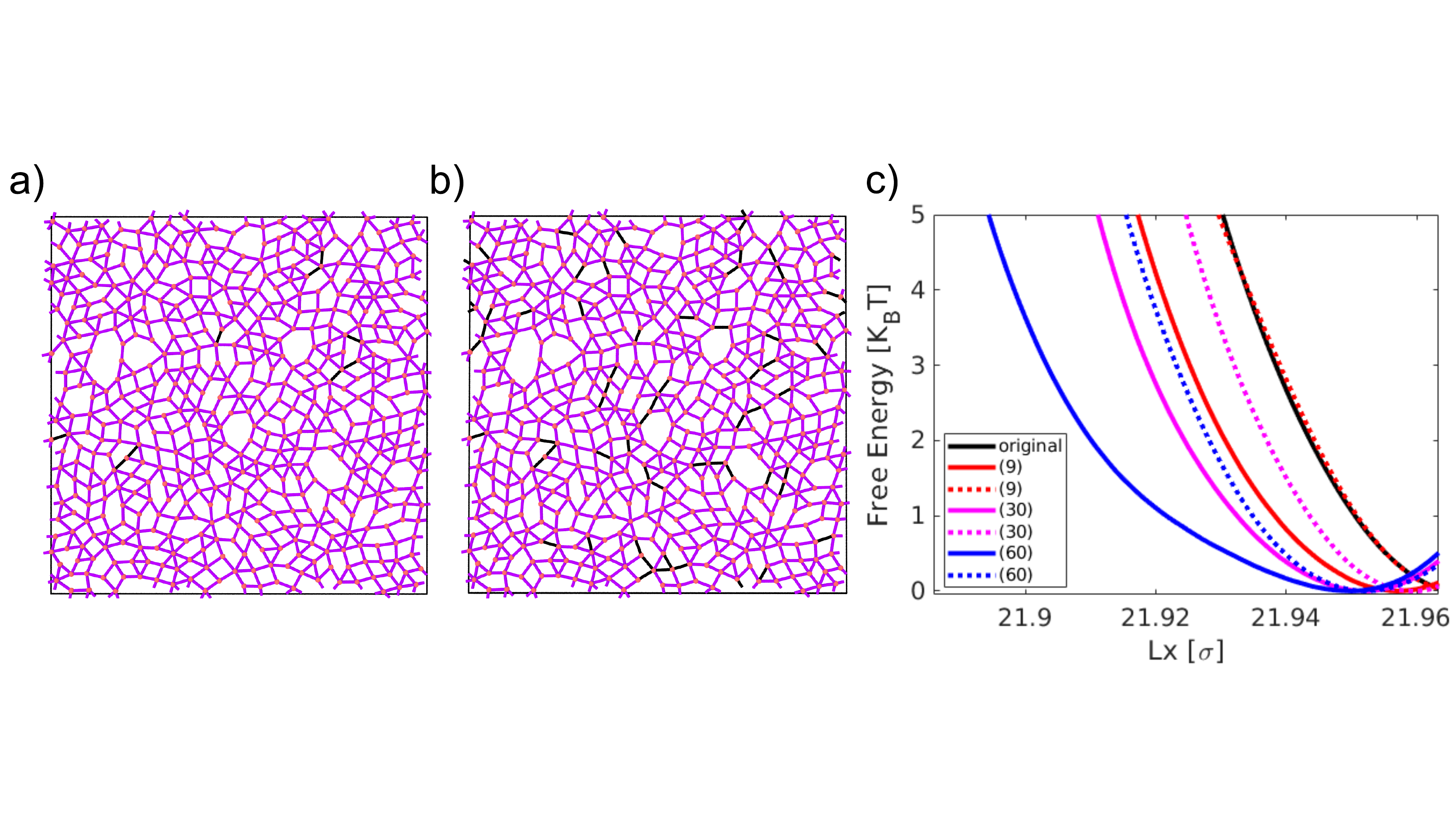}   
    \caption{(a) Illustration of the simulated network with the top 9 ($\sim 1\%$) HLDA bonds highlighted in dark blue, and (b) with the top 60 ($\sim 6\%$) HLDA bonds highlighted in dark blue. (c) The computed FES of the simulated pristine network  as function of $L_x$ (solid black) along with the FES of modified realizations of the network in which the bond coefficient of the highest weighted HLDA bonds was reduced to $0.01 \epsilon / \sigma^2$ (solid lines) or in which the bond coefficient of randomly selected bonds was reduced to $0.01 \epsilon / \sigma^2$ (doted lines). In parenthesis, the percentage of bonds that was modified in the network.}
    \label{fig:Compression results}
\end{figure}

Defining the 1007 distances corresponding to all of the network's bonds as our descriptor set and calculating the relaxed and compressed states' expectation vectors and covariance matrices, we apply Eq. \ref{maximizer} to obtain the HLDA CV. As previously done, to circumvent the limitation imposed by the linearity of HLDA, we apply it iteratively, whereby the descriptors attaining the three largest weights in absolute value at each iteration are selected and removed from the descriptor set used in the iterations to follow. Figs. \ref{fig:Compression results}a and \ref{fig:Compression results}b highlight, respectively, the top $\sim 1 \%$ and top $\sim 6 \%$ of bonds selected in this manner. The selected bonds are distributed fairly homogeneously across the network, in contrast to the allosteric case. Such bonds appear to be organized in small clusters, consisting of 2-6 bonds each. 

To modify the FES corresponding to the network's uniaxial strain fluctuations, we systematically alter the bond coefficient of the selected bonds. To quantify the resulting behavior, we simulate the network under constant temperature and constant pressure, applying the Parrinello-Rahman barostat \cite{parrinello1981polymorphic,martyna1994constant} in the $x$ direction of the simulation cell, keeping the simulation cell edge in the perpendicular direction, $L_y$, constant.  From these simulations, we compute the probability density $P(L_x)$ and, using Eq. \ref{collective variables defined}, $F(Lx)$, the corresponding FES of the network \cite{martovnak2003predicting}.  

Fig. \ref{fig:Compression results}b exhibits the FES of the pristine and modified networks. Altering the bonds selected by CV-FEST gives rise to substantially greater changes in the system's FES compared to the case in which randomly selected bonds are altered. This is particularly apparent when less than $1\%$ (i.e. 9 bonds) of the network's bonds are altered. In that case, CV-FEST is able to select strategically important bonds of the network, whereas a random selection yields no apparent change of the system's FES. This stark difference illustrates the importance of the critical bonds' positions in the network, given that all the bonds in the pristine network possess the same elastic modulus. Examining the FES of the different networks, we find that altering the bond coefficient of the selected bonds leads to three different effects. The first is a modification of the steepness of the FES. The second, is a change of the functional form of the FES, namely the extent to which it deviates from a parabola (as would be expected for entropic contributions to the system's elasticity). The third is a shift of the minimum. The ability to modify the FES of the network in these ways points to the promise of using CV-FEST to engineer systems that exhibit target, desirable mechanical properties.


\section*{Methods}

\subsubsection*{Network Construction}
Construction of the networks followed the protocol put forward in ref. \cite{rocks2017designing}. Briefly, 2D  configurations of soft disks placed in a simulation cell with periodic boundary conditions, and allowed to relax to a local energy minimum using a standard jamming algorithm \cite{liu2010jamming}.  The network is then constructed by  placing  nodes  at  the  center of each disk, and by linking nodes corresponding to disks which overlap in the resulting configuration. To implement the elastic networks, beads of identical mass are placed at every node position in the network, and links are replaced by harmonic bonds with an elastic energy of the form:

\begin{equation}
\label{bond_energy}
V(\mathbf{r}_{ij})=\frac{K_{ij}}{(\mathbf{r}^0_{ij})^2}\left(\mathbf{r}_{ij}-\mathbf{r}^0_{ij} \right)^2
\end{equation}

\noindent where $K_{ij}$  is the bond coefficient corresponding to the bond between the $i$ and $j$ beads and is set initially to 1 for all bonds in the network; $\mathbf{r}^0_{ij}$ is the rest length of the bond, and $\mathbf{r}_{ij}$ is the distance between the beads.

\subsubsection*{Embedding allosteric response into the network}
Allosteric response is embedded into the network by randomly selecting two pairs of neighboring non-bonded beads, referred to as the source beads and target beads (see Fig. 1). The target beads are chosen to be spatially distant from the source beads to achieve the long-range effect that characterizes allostery.

The allosteric effect is defined as an imposed change of the distance between the target beads given a change in the distance between the source beads. To optimize the network such that this effect will emerge, a tuning-by-pruning of bonds strategy is employed with the objective to minimize the fitness function of Eq. \ref{allosteric_optimization}, which measures the difference between the desired target beads' response and the actual response \cite{rocks2017designing,rocks2019limits}. Namely, the ratio of the target strain to the source strain, $\eta = \Delta d_t / \Delta d_s$, is measured and compared to the desired ratio $\eta^*$, set to 5, rendering the fitness function to be:

\begin{equation}
\label{allosteric_optimization}
\Delta^2=(\eta/\eta^* -1)^2.
\end{equation}

\noindent To compute Eq. \ref{allosteric_optimization}, the source beads are “pinched” to $50 \%$ of their initial distance and frozen at their new positions, after which a second minimization of network energy is carried out and the ratio $\eta$ is calculated. The optimization procedure was applied iteratively, whereby at each iteration,  $\Delta^2$ resulting from a trial removal of each bond in the network was computed. A greedy algorithm was followed in which the bond the removal of which lead to the largest decrease in $\Delta^2$ with respect to its previous value, was permanently deleted. To keep local stability, however, the bond is deleted only if all the beads it was connected to were connected to at least three remaining bonds \cite{goodrich2015principle}. Otherwise, the bond that created the next-largest decrease in $\Delta^2$ is permanently deleted, given that it satisfied this constraint, and so on. This iterative process is continued until the desired strain ratio (Eq. \ref{allosteric_optimization}) is attained. The energy minimization of the network was performed using LAMMPS \cite{PLIMPTON1995}.

\subsubsection*{Dynamic simulations of the Allosteric network}
All simulations were run with LAMMPS \cite{PLIMPTON1995} patched with PLUMED2.6 \cite{tribello_2014}. The network consisted of 499 beads and 999 bonds. For simulations run at $T=8.6 \cdot 10^{-6}$ all beads had a mass of $M=m$ with the exception of the source beads which had a mass of $M_s=1000m$. For the sake of preventing the destabilization of the network, for WTMD simulations run at $T=4.3 \cdot 10^{-5}$ all beads had a mass of $M=100m$ with the exception of the source beads which had a mass of $M_s=1000m$.  All simulations were initially energetically relaxed at zero pressure and subsequently run at a constant temperature using a Langevin thermostat \cite{schneider1978molecular} with a damping parameter of 1, and a time step of $0.001 \tau$. In the unbiased training simulations, the source bead positions were constrained in the allosteric activated state B, $d_s=1.87 \sigma$, and the trap state C, $d_s=2.1 \sigma$, to keep the network from relaxing back to the inactivated state A. The free energy surface of the system was computed using Eq.\ref{Eq: WTMD}:

\begin{equation}
\label{Eq: WTMD}
F(\mathbf{s})=\frac{\gamma}{1 - \gamma}V(\mathbf{s})
\end{equation}

\noindent where $V(\mathbf{s})$ is the bias potential deposited in the WTMD simulations, and $\gamma$ is the so called bias factor. WTMD simulations were run with a bias factor of $\gamma=180$, hill height of $0.0001 \epsilon$, and hill width of $0.01 \sigma$. 

\subsubsection*{Probing uniaxial fluctuations}
Simulations were run with LAMMPS \cite{PLIMPTON1995} patched with PLUMED2.6 \cite{tribello_2014}. The network consisted of 499 beads and 1007 bonds. All beads had a mass m. Training simulations, in which the networks were slightly compressed in the x direction, were run at constant temperature $T=8.6 \cdot 10^{-6}$ (with $k_B=1$) using a Langevin thermostat \cite{schneider1978molecular} with a damping parameter of 1, and a time step of $0.01 \tau$. Compression was executed using a constant deformation rate of $10^{-7} \tau^{-1}$. The FESs of the networks were measured in constant pressure simulations run at zero pressure, using a Parrinello-Rahman barostat \cite{martyna1994constant,parrinello1980crystal} applied in the x direction, corresponding to the direction of compression in the training simulations. The simulation box side in the y direction was kept constant in all simulations.

\section*{Conclusions}

In conclusion, we have studied the ability to systematically tailor the FES of a disordered elastic network with respect to allosteric regulation and uniaxial strain fluctuations at finite temperature and pressure conditions. We find that CV-FEST is capable of (1) identifying the important bonds in the network with respect to each of these functionalities and, (2) by altering these bonds' stiffness, of tailoring in a tractable way the system's FES and the corresponding functional behavior. 

Given the complex interconnected nature of the networks, CV-FEST's demonstrated capabilities offer potential as a tool for design and analysis of complex systems in general, including systems such as proteins, macromolecules, and materials. Finally, considering that CV-FEST relies solely on kinematic information for its input, it would be interesting to explore its direct applicability to experimental systems for which such information may be relatively easily generated. 

\section*{Acknowledgement}
This work is supported by the Department of Energy, Basic Energy Sciences, through the Midwest Center for Computational Materials (MiCCoM).


\newpage
\def\bibsection{\section*{\refname}} 

\bibliography{library}

\begin{thebibliography}{46}%
\makeatletter
\providecommand \@ifxundefined [1]{%
 \@ifx{#1\undefined}
}%
\providecommand \@ifnum [1]{%
 \ifnum #1\expandafter \@firstoftwo
 \else \expandafter \@secondoftwo
 \fi
}%
\providecommand \@ifx [1]{%
 \ifx #1\expandafter \@firstoftwo
 \else \expandafter \@secondoftwo
 \fi
}%
\providecommand \natexlab [1]{#1}%
\providecommand \enquote  [1]{``#1''}%
\providecommand \bibnamefont  [1]{#1}%
\providecommand \bibfnamefont [1]{#1}%
\providecommand \citenamefont [1]{#1}%
\providecommand \href@noop [0]{\@secondoftwo}%
\providecommand \href [0]{\begingroup \@sanitize@url \@href}%
\providecommand \@href[1]{\@@startlink{#1}\@@href}%
\providecommand \@@href[1]{\endgroup#1\@@endlink}%
\providecommand \@sanitize@url [0]{\catcode `\\12\catcode `\$12\catcode
  `\&12\catcode `\#12\catcode `\^12\catcode `\_12\catcode `\%12\relax}%
\providecommand \@@startlink[1]{}%
\providecommand \@@endlink[0]{}%
\providecommand \url  [0]{\begingroup\@sanitize@url \@url }%
\providecommand \@url [1]{\endgroup\@href {#1}{\urlprefix }}%
\providecommand \urlprefix  [0]{URL }%
\providecommand \Eprint [0]{\href }%
\providecommand \doibase [0]{https://doi.org/}%
\providecommand \selectlanguage [0]{\@gobble}%
\providecommand \bibinfo  [0]{\@secondoftwo}%
\providecommand \bibfield  [0]{\@secondoftwo}%
\providecommand \translation [1]{[#1]}%
\providecommand \BibitemOpen [0]{}%
\providecommand \bibitemStop [0]{}%
\providecommand \bibitemNoStop [0]{.\EOS\space}%
\providecommand \EOS [0]{\spacefactor3000\relax}%
\providecommand \BibitemShut  [1]{\csname bibitem#1\endcsname}%
\let\auto@bib@innerbib\@empty
\bibitem [{\citenamefont {Long}\ and\ \citenamefont
  {Ferguson}(2018)}]{long2018rational}%
  \BibitemOpen
  \bibfield  {author} {\bibinfo {author} {\bibfnamefont {A.~W.}\ \bibnamefont
  {Long}}\ and\ \bibinfo {author} {\bibfnamefont {A.~L.}\ \bibnamefont
  {Ferguson}},\ }\bibfield  {title} {\bibinfo {title} {Rational design of
  patchy colloids via landscape engineering},\ }\href@noop {} {\bibfield
  {journal} {\bibinfo  {journal} {Molecular Systems Design \& Engineering}\
  }\textbf {\bibinfo {volume} {3}},\ \bibinfo {pages} {49} (\bibinfo {year}
  {2018})}\BibitemShut {NoStop}%
\bibitem [{\citenamefont {White}\ and\ \citenamefont
  {Voth}(2014)}]{white2014efficient}%
  \BibitemOpen
  \bibfield  {author} {\bibinfo {author} {\bibfnamefont {A.~D.}\ \bibnamefont
  {White}}\ and\ \bibinfo {author} {\bibfnamefont {G.~A.}\ \bibnamefont
  {Voth}},\ }\bibfield  {title} {\bibinfo {title} {Efficient and minimal method
  to bias molecular simulations with experimental data},\ }\href@noop {}
  {\bibfield  {journal} {\bibinfo  {journal} {Journal of chemical theory and
  computation}\ }\textbf {\bibinfo {volume} {10}},\ \bibinfo {pages} {3023}
  (\bibinfo {year} {2014})}\BibitemShut {NoStop}%
\bibitem [{\citenamefont {Amirkulova}\ and\ \citenamefont
  {White}(2019)}]{amirkulova2019recent}%
  \BibitemOpen
  \bibfield  {author} {\bibinfo {author} {\bibfnamefont {D.~B.}\ \bibnamefont
  {Amirkulova}}\ and\ \bibinfo {author} {\bibfnamefont {A.~D.}\ \bibnamefont
  {White}},\ }\bibfield  {title} {\bibinfo {title} {Recent advances in maximum
  entropy biasing techniques for molecular dynamics},\ }\href@noop {}
  {\bibfield  {journal} {\bibinfo  {journal} {Molecular Simulation}\ }\textbf
  {\bibinfo {volume} {45}},\ \bibinfo {pages} {1285} (\bibinfo {year}
  {2019})}\BibitemShut {NoStop}%
\bibitem [{\citenamefont {Gil-Ley}\ \emph {et~al.}(2016)\citenamefont
  {Gil-Ley}, \citenamefont {Bottaro},\ and\ \citenamefont
  {Bussi}}]{gil2016empirical}%
  \BibitemOpen
  \bibfield  {author} {\bibinfo {author} {\bibfnamefont {A.}~\bibnamefont
  {Gil-Ley}}, \bibinfo {author} {\bibfnamefont {S.}~\bibnamefont {Bottaro}},\
  and\ \bibinfo {author} {\bibfnamefont {G.}~\bibnamefont {Bussi}},\ }\bibfield
   {title} {\bibinfo {title} {Empirical corrections to the amber rna force
  field with target metadynamics},\ }\href@noop {} {\bibfield  {journal}
  {\bibinfo  {journal} {Journal of chemical theory and computation}\ }\textbf
  {\bibinfo {volume} {12}},\ \bibinfo {pages} {2790} (\bibinfo {year}
  {2016})}\BibitemShut {NoStop}%
\bibitem [{\citenamefont {Marinelli}\ and\ \citenamefont
  {Faraldo-G{\'o}mez}(2015)}]{marinelli2015ensemble}%
  \BibitemOpen
  \bibfield  {author} {\bibinfo {author} {\bibfnamefont {F.}~\bibnamefont
  {Marinelli}}\ and\ \bibinfo {author} {\bibfnamefont {J.~D.}\ \bibnamefont
  {Faraldo-G{\'o}mez}},\ }\bibfield  {title} {\bibinfo {title} {Ensemble-biased
  metadynamics: a molecular simulation method to sample experimental
  distributions},\ }\href@noop {} {\bibfield  {journal} {\bibinfo  {journal}
  {Biophysical journal}\ }\textbf {\bibinfo {volume} {108}},\ \bibinfo {pages}
  {2779} (\bibinfo {year} {2015})}\BibitemShut {NoStop}%
\bibitem [{\citenamefont {Cesari}\ \emph {et~al.}(2018)\citenamefont {Cesari},
  \citenamefont {Rei{\ss}er},\ and\ \citenamefont {Bussi}}]{cesari2018using}%
  \BibitemOpen
  \bibfield  {author} {\bibinfo {author} {\bibfnamefont {A.}~\bibnamefont
  {Cesari}}, \bibinfo {author} {\bibfnamefont {S.}~\bibnamefont {Rei{\ss}er}},\
  and\ \bibinfo {author} {\bibfnamefont {G.}~\bibnamefont {Bussi}},\ }\bibfield
   {title} {\bibinfo {title} {Using the maximum entropy principle to combine
  simulations and solution experiments},\ }\href@noop {} {\bibfield  {journal}
  {\bibinfo  {journal} {Computation}\ }\textbf {\bibinfo {volume} {6}},\
  \bibinfo {pages} {15} (\bibinfo {year} {2018})}\BibitemShut {NoStop}%
\bibitem [{\citenamefont {Mendels}\ and\ \citenamefont
  {de~Pablo}(2022)}]{mendels2022collective}%
  \BibitemOpen
  \bibfield  {author} {\bibinfo {author} {\bibfnamefont {D.}~\bibnamefont
  {Mendels}}\ and\ \bibinfo {author} {\bibfnamefont {J.~J.}\ \bibnamefont
  {de~Pablo}},\ }\bibfield  {title} {\bibinfo {title} {Collective variables for
  free energy surface tailoring: Understanding and modifying functionality in
  systems dominated by rare events},\ }\href@noop {} {\bibfield  {journal}
  {\bibinfo  {journal} {The Journal of Physical Chemistry Letters}\ }\textbf
  {\bibinfo {volume} {13}},\ \bibinfo {pages} {2830} (\bibinfo {year}
  {2022})}\BibitemShut {NoStop}%
\bibitem [{\citenamefont {Bahar}\ \emph {et~al.}(2010)\citenamefont {Bahar},
  \citenamefont {Lezon}, \citenamefont {Yang},\ and\ \citenamefont
  {Eyal}}]{bahar2010global}%
  \BibitemOpen
  \bibfield  {author} {\bibinfo {author} {\bibfnamefont {I.}~\bibnamefont
  {Bahar}}, \bibinfo {author} {\bibfnamefont {T.~R.}\ \bibnamefont {Lezon}},
  \bibinfo {author} {\bibfnamefont {L.-W.}\ \bibnamefont {Yang}},\ and\
  \bibinfo {author} {\bibfnamefont {E.}~\bibnamefont {Eyal}},\ }\bibfield
  {title} {\bibinfo {title} {Global dynamics of proteins: bridging between
  structure and function},\ }\href@noop {} {\bibfield  {journal} {\bibinfo
  {journal} {Annual review of biophysics}\ }\textbf {\bibinfo {volume} {39}},\
  \bibinfo {pages} {23} (\bibinfo {year} {2010})}\BibitemShut {NoStop}%
\bibitem [{\citenamefont {Rocks}\ \emph {et~al.}(2017)\citenamefont {Rocks},
  \citenamefont {Pashine}, \citenamefont {Bischofberger}, \citenamefont
  {Goodrich}, \citenamefont {Liu},\ and\ \citenamefont
  {Nagel}}]{rocks2017designing}%
  \BibitemOpen
  \bibfield  {author} {\bibinfo {author} {\bibfnamefont {J.~W.}\ \bibnamefont
  {Rocks}}, \bibinfo {author} {\bibfnamefont {N.}~\bibnamefont {Pashine}},
  \bibinfo {author} {\bibfnamefont {I.}~\bibnamefont {Bischofberger}}, \bibinfo
  {author} {\bibfnamefont {C.~P.}\ \bibnamefont {Goodrich}}, \bibinfo {author}
  {\bibfnamefont {A.~J.}\ \bibnamefont {Liu}},\ and\ \bibinfo {author}
  {\bibfnamefont {S.~R.}\ \bibnamefont {Nagel}},\ }\bibfield  {title} {\bibinfo
  {title} {Designing allostery-inspired response in mechanical networks},\
  }\href@noop {} {\bibfield  {journal} {\bibinfo  {journal} {Proceedings of the
  National Academy of Sciences}\ }\textbf {\bibinfo {volume} {114}},\ \bibinfo
  {pages} {2520} (\bibinfo {year} {2017})}\BibitemShut {NoStop}%
\bibitem [{\citenamefont {Reid}\ \emph {et~al.}(2018)\citenamefont {Reid},
  \citenamefont {Pashine}, \citenamefont {Wozniak}, \citenamefont {Jaeger},
  \citenamefont {Liu}, \citenamefont {Nagel},\ and\ \citenamefont
  {de~Pablo}}]{reid2018auxetic}%
  \BibitemOpen
  \bibfield  {author} {\bibinfo {author} {\bibfnamefont {D.~R.}\ \bibnamefont
  {Reid}}, \bibinfo {author} {\bibfnamefont {N.}~\bibnamefont {Pashine}},
  \bibinfo {author} {\bibfnamefont {J.~M.}\ \bibnamefont {Wozniak}}, \bibinfo
  {author} {\bibfnamefont {H.~M.}\ \bibnamefont {Jaeger}}, \bibinfo {author}
  {\bibfnamefont {A.~J.}\ \bibnamefont {Liu}}, \bibinfo {author} {\bibfnamefont
  {S.~R.}\ \bibnamefont {Nagel}},\ and\ \bibinfo {author} {\bibfnamefont
  {J.~J.}\ \bibnamefont {de~Pablo}},\ }\bibfield  {title} {\bibinfo {title}
  {Auxetic metamaterials from disordered networks},\ }\href@noop {} {\bibfield
  {journal} {\bibinfo  {journal} {Proceedings of the National Academy of
  Sciences}\ }\textbf {\bibinfo {volume} {115}},\ \bibinfo {pages} {E1384}
  (\bibinfo {year} {2018})}\BibitemShut {NoStop}%
\bibitem [{\citenamefont {Hexner}\ \emph {et~al.}(2020)\citenamefont {Hexner},
  \citenamefont {Liu},\ and\ \citenamefont {Nagel}}]{hexner2020periodic}%
  \BibitemOpen
  \bibfield  {author} {\bibinfo {author} {\bibfnamefont {D.}~\bibnamefont
  {Hexner}}, \bibinfo {author} {\bibfnamefont {A.~J.}\ \bibnamefont {Liu}},\
  and\ \bibinfo {author} {\bibfnamefont {S.~R.}\ \bibnamefont {Nagel}},\
  }\bibfield  {title} {\bibinfo {title} {Periodic training of creeping
  solids},\ }\href@noop {} {\bibfield  {journal} {\bibinfo  {journal}
  {Proceedings of the National Academy of Sciences}\ }\textbf {\bibinfo
  {volume} {117}},\ \bibinfo {pages} {31690} (\bibinfo {year}
  {2020})}\BibitemShut {NoStop}%
\bibitem [{\citenamefont {Albert}\ and\ \citenamefont
  {Barab{\'a}si}(2002)}]{albert2002statistical}%
  \BibitemOpen
  \bibfield  {author} {\bibinfo {author} {\bibfnamefont {R.}~\bibnamefont
  {Albert}}\ and\ \bibinfo {author} {\bibfnamefont {A.-L.}\ \bibnamefont
  {Barab{\'a}si}},\ }\bibfield  {title} {\bibinfo {title} {Statistical
  mechanics of complex networks},\ }\href@noop {} {\bibfield  {journal}
  {\bibinfo  {journal} {Reviews of modern physics}\ }\textbf {\bibinfo {volume}
  {74}},\ \bibinfo {pages} {47} (\bibinfo {year} {2002})}\BibitemShut {NoStop}%
\bibitem [{\citenamefont {Liu}\ and\ \citenamefont
  {Barab{\'a}si}(2016)}]{liu2016control}%
  \BibitemOpen
  \bibfield  {author} {\bibinfo {author} {\bibfnamefont {Y.-Y.}\ \bibnamefont
  {Liu}}\ and\ \bibinfo {author} {\bibfnamefont {A.-L.}\ \bibnamefont
  {Barab{\'a}si}},\ }\bibfield  {title} {\bibinfo {title} {Control principles
  of complex systems},\ }\href@noop {} {\bibfield  {journal} {\bibinfo
  {journal} {Reviews of Modern Physics}\ }\textbf {\bibinfo {volume} {88}},\
  \bibinfo {pages} {035006} (\bibinfo {year} {2016})}\BibitemShut {NoStop}%
\bibitem [{\citenamefont {Torrie}\ and\ \citenamefont
  {Valleau}(1977)}]{Torrie1977}%
  \BibitemOpen
  \bibfield  {author} {\bibinfo {author} {\bibfnamefont {G.~M.}\ \bibnamefont
  {Torrie}}\ and\ \bibinfo {author} {\bibfnamefont {J.~P.}\ \bibnamefont
  {Valleau}},\ }\bibfield  {title} {\bibinfo {title} {{Nonphysical sampling
  distributions in Monte Carlo free-energy estimation: Umbrella sampling}},\
  }\href {https://doi.org/10.1016/0021-9991(77)90121-8} {\bibfield  {journal}
  {\bibinfo  {journal} {J. Comput. Phys.}\ }\textbf {\bibinfo {volume} {23}},\
  \bibinfo {pages} {187} (\bibinfo {year} {1977})}\BibitemShut {NoStop}%
\bibitem [{\citenamefont {Laio}\ and\ \citenamefont
  {Parrinello}(2002)}]{laio_parrinello_2002}%
  \BibitemOpen
  \bibfield  {author} {\bibinfo {author} {\bibfnamefont {A.}~\bibnamefont
  {Laio}}\ and\ \bibinfo {author} {\bibfnamefont {M.}~\bibnamefont
  {Parrinello}},\ }\bibfield  {title} {\bibinfo {title} {{Escaping free-energy
  minima}},\ }\href {https://doi.org/10.1073/pnas.202427399} {\bibfield
  {journal} {\bibinfo  {journal} {Proc. Natl. Acad. Sci. U.S.A.}\ }\textbf
  {\bibinfo {volume} {99}},\ \bibinfo {pages} {12562} (\bibinfo {year}
  {2002})}\BibitemShut {NoStop}%
\bibitem [{\citenamefont {Darve}\ and\ \citenamefont
  {Pohorille}(2001)}]{darve2001calculating}%
  \BibitemOpen
  \bibfield  {author} {\bibinfo {author} {\bibfnamefont {E.}~\bibnamefont
  {Darve}}\ and\ \bibinfo {author} {\bibfnamefont {A.}~\bibnamefont
  {Pohorille}},\ }\bibfield  {title} {\bibinfo {title} {Calculating free
  energies using average force},\ }\href@noop {} {\bibfield  {journal}
  {\bibinfo  {journal} {J. Chem. Phys.}\ }\textbf {\bibinfo {volume} {115}},\
  \bibinfo {pages} {9169} (\bibinfo {year} {2001})}\BibitemShut {NoStop}%
\bibitem [{\citenamefont {Ribeiro}\ \emph {et~al.}(2018)\citenamefont
  {Ribeiro}, \citenamefont {Bravo}, \citenamefont {Wang},\ and\ \citenamefont
  {Tiwary}}]{ribeiro2018reweighted}%
  \BibitemOpen
  \bibfield  {author} {\bibinfo {author} {\bibfnamefont {J.~M.~L.}\
  \bibnamefont {Ribeiro}}, \bibinfo {author} {\bibfnamefont {P.}~\bibnamefont
  {Bravo}}, \bibinfo {author} {\bibfnamefont {Y.}~\bibnamefont {Wang}},\ and\
  \bibinfo {author} {\bibfnamefont {P.}~\bibnamefont {Tiwary}},\ }\bibfield
  {title} {\bibinfo {title} {Reweighted autoencoded variational bayes for
  enhanced sampling (rave)},\ }\href@noop {} {\bibfield  {journal} {\bibinfo
  {journal} {The Journal of chemical physics}\ }\textbf {\bibinfo {volume}
  {149}},\ \bibinfo {pages} {072301} (\bibinfo {year} {2018})}\BibitemShut
  {NoStop}%
\bibitem [{\citenamefont {Mendels}\ \emph
  {et~al.}(2018{\natexlab{a}})\citenamefont {Mendels}, \citenamefont
  {Piccini},\ and\ \citenamefont {Parrinello}}]{mendels2018collective}%
  \BibitemOpen
  \bibfield  {author} {\bibinfo {author} {\bibfnamefont {D.}~\bibnamefont
  {Mendels}}, \bibinfo {author} {\bibfnamefont {G.}~\bibnamefont {Piccini}},\
  and\ \bibinfo {author} {\bibfnamefont {M.}~\bibnamefont {Parrinello}},\
  }\bibfield  {title} {\bibinfo {title} {Collective variables from local
  fluctuations},\ }\href@noop {} {\bibfield  {journal} {\bibinfo  {journal} {J.
  Phys. Chem. Lett.}\ }\textbf {\bibinfo {volume} {9}},\ \bibinfo {pages}
  {2776} (\bibinfo {year} {2018}{\natexlab{a}})}\BibitemShut {NoStop}%
\bibitem [{\citenamefont {Bonati}\ \emph {et~al.}(2020)\citenamefont {Bonati},
  \citenamefont {Rizzi},\ and\ \citenamefont {Parrinello}}]{bonati2020data}%
  \BibitemOpen
  \bibfield  {author} {\bibinfo {author} {\bibfnamefont {L.}~\bibnamefont
  {Bonati}}, \bibinfo {author} {\bibfnamefont {V.}~\bibnamefont {Rizzi}},\ and\
  \bibinfo {author} {\bibfnamefont {M.}~\bibnamefont {Parrinello}},\ }\bibfield
   {title} {\bibinfo {title} {Data-driven collective variables for enhanced
  sampling},\ }\href@noop {} {\bibfield  {journal} {\bibinfo  {journal} {The
  journal of physical chemistry letters}\ }\textbf {\bibinfo {volume} {11}},\
  \bibinfo {pages} {2998} (\bibinfo {year} {2020})}\BibitemShut {NoStop}%
\bibitem [{\citenamefont {Chen}\ \emph {et~al.}(2018)\citenamefont {Chen},
  \citenamefont {Tan},\ and\ \citenamefont {Ferguson}}]{chen2018collective}%
  \BibitemOpen
  \bibfield  {author} {\bibinfo {author} {\bibfnamefont {W.}~\bibnamefont
  {Chen}}, \bibinfo {author} {\bibfnamefont {A.~R.}\ \bibnamefont {Tan}},\ and\
  \bibinfo {author} {\bibfnamefont {A.~L.}\ \bibnamefont {Ferguson}},\
  }\bibfield  {title} {\bibinfo {title} {Collective variable discovery and
  enhanced sampling using autoencoders: Innovations in network architecture and
  error function design},\ }\href@noop {} {\bibfield  {journal} {\bibinfo
  {journal} {The Journal of chemical physics}\ }\textbf {\bibinfo {volume}
  {149}},\ \bibinfo {pages} {072312} (\bibinfo {year} {2018})}\BibitemShut
  {NoStop}%
\bibitem [{\citenamefont {Wehmeyer}\ and\ \citenamefont
  {No{\'e}}(2018)}]{wehmeyer2018time}%
  \BibitemOpen
  \bibfield  {author} {\bibinfo {author} {\bibfnamefont {C.}~\bibnamefont
  {Wehmeyer}}\ and\ \bibinfo {author} {\bibfnamefont {F.}~\bibnamefont
  {No{\'e}}},\ }\bibfield  {title} {\bibinfo {title} {Time-lagged autoencoders:
  Deep learning of slow collective variables for molecular kinetics},\
  }\href@noop {} {\bibfield  {journal} {\bibinfo  {journal} {The Journal of
  chemical physics}\ }\textbf {\bibinfo {volume} {148}},\ \bibinfo {pages}
  {241703} (\bibinfo {year} {2018})}\BibitemShut {NoStop}%
\bibitem [{\citenamefont {Sultan}\ and\ \citenamefont
  {Pande}(2018)}]{sultan2018automated}%
  \BibitemOpen
  \bibfield  {author} {\bibinfo {author} {\bibfnamefont {M.~M.}\ \bibnamefont
  {Sultan}}\ and\ \bibinfo {author} {\bibfnamefont {V.~S.}\ \bibnamefont
  {Pande}},\ }\bibfield  {title} {\bibinfo {title} {Automated design of
  collective variables using supervised machine learning},\ }\href@noop {}
  {\bibfield  {journal} {\bibinfo  {journal} {The Journal of chemical physics}\
  }\textbf {\bibinfo {volume} {149}},\ \bibinfo {pages} {094106} (\bibinfo
  {year} {2018})}\BibitemShut {NoStop}%
\bibitem [{\citenamefont {McCarty}\ and\ \citenamefont
  {Parrinello}(2017)}]{mccarty2017variational}%
  \BibitemOpen
  \bibfield  {author} {\bibinfo {author} {\bibfnamefont {J.}~\bibnamefont
  {McCarty}}\ and\ \bibinfo {author} {\bibfnamefont {M.}~\bibnamefont
  {Parrinello}},\ }\bibfield  {title} {\bibinfo {title} {A variational
  conformational dynamics approach to the selection of collective variables in
  metadynamics},\ }\href@noop {} {\bibfield  {journal} {\bibinfo  {journal} {J.
  Chem. Phys.}\ }\textbf {\bibinfo {volume} {147}},\ \bibinfo {pages} {204109}
  (\bibinfo {year} {2017})}\BibitemShut {NoStop}%
\bibitem [{\citenamefont {Piccini}\ \emph {et~al.}(2018)\citenamefont
  {Piccini}, \citenamefont {Mendels},\ and\ \citenamefont
  {Parrinello}}]{piccini2018metadynamics}%
  \BibitemOpen
  \bibfield  {author} {\bibinfo {author} {\bibfnamefont {G.}~\bibnamefont
  {Piccini}}, \bibinfo {author} {\bibfnamefont {D.}~\bibnamefont {Mendels}},\
  and\ \bibinfo {author} {\bibfnamefont {M.}~\bibnamefont {Parrinello}},\
  }\bibfield  {title} {\bibinfo {title} {Metadynamics with discriminants: A
  tool for understanding chemistry},\ }\href@noop {} {\bibfield  {journal}
  {\bibinfo  {journal} {Journal of chemical theory and computation}\ }\textbf
  {\bibinfo {volume} {14}},\ \bibinfo {pages} {5040} (\bibinfo {year}
  {2018})}\BibitemShut {NoStop}%
\bibitem [{\citenamefont {Mendels}\ \emph
  {et~al.}(2018{\natexlab{b}})\citenamefont {Mendels}, \citenamefont {Piccini},
  \citenamefont {Brotzakis}, \citenamefont {Yang},\ and\ \citenamefont
  {Parrinello}}]{mendels2018folding}%
  \BibitemOpen
  \bibfield  {author} {\bibinfo {author} {\bibfnamefont {D.}~\bibnamefont
  {Mendels}}, \bibinfo {author} {\bibfnamefont {G.}~\bibnamefont {Piccini}},
  \bibinfo {author} {\bibfnamefont {Z.~F.}\ \bibnamefont {Brotzakis}}, \bibinfo
  {author} {\bibfnamefont {Y.~I.}\ \bibnamefont {Yang}},\ and\ \bibinfo
  {author} {\bibfnamefont {M.}~\bibnamefont {Parrinello}},\ }\bibfield  {title}
  {\bibinfo {title} {Folding a small protein using harmonic linear discriminant
  analysis},\ }\href@noop {} {\bibfield  {journal} {\bibinfo  {journal} {The
  Journal of chemical physics}\ }\textbf {\bibinfo {volume} {149}},\ \bibinfo
  {pages} {194113} (\bibinfo {year} {2018}{\natexlab{b}})}\BibitemShut
  {NoStop}%
\bibitem [{\citenamefont {Rizzi}\ \emph {et~al.}(2019)\citenamefont {Rizzi},
  \citenamefont {Mendels}, \citenamefont {Sicilia},\ and\ \citenamefont
  {Parrinello}}]{rizzi2019blind}%
  \BibitemOpen
  \bibfield  {author} {\bibinfo {author} {\bibfnamefont {V.}~\bibnamefont
  {Rizzi}}, \bibinfo {author} {\bibfnamefont {D.}~\bibnamefont {Mendels}},
  \bibinfo {author} {\bibfnamefont {E.}~\bibnamefont {Sicilia}},\ and\ \bibinfo
  {author} {\bibfnamefont {M.}~\bibnamefont {Parrinello}},\ }\bibfield  {title}
  {\bibinfo {title} {Blind search for complex chemical pathways using harmonic
  linear discriminant analysis},\ }\href@noop {} {\bibfield  {journal}
  {\bibinfo  {journal} {Journal of chemical theory and computation}\ }\textbf
  {\bibinfo {volume} {15}},\ \bibinfo {pages} {4507} (\bibinfo {year}
  {2019})}\BibitemShut {NoStop}%
\bibitem [{\citenamefont {Zhang}\ \emph {et~al.}(2019)\citenamefont {Zhang},
  \citenamefont {Niu}, \citenamefont {Piccini}, \citenamefont {Mendels},\ and\
  \citenamefont {Parrinello}}]{zhang2019improving}%
  \BibitemOpen
  \bibfield  {author} {\bibinfo {author} {\bibfnamefont {Y.-Y.}\ \bibnamefont
  {Zhang}}, \bibinfo {author} {\bibfnamefont {H.}~\bibnamefont {Niu}}, \bibinfo
  {author} {\bibfnamefont {G.}~\bibnamefont {Piccini}}, \bibinfo {author}
  {\bibfnamefont {D.}~\bibnamefont {Mendels}},\ and\ \bibinfo {author}
  {\bibfnamefont {M.}~\bibnamefont {Parrinello}},\ }\bibfield  {title}
  {\bibinfo {title} {Improving collective variables: The case of
  crystallization},\ }\href@noop {} {\bibfield  {journal} {\bibinfo  {journal}
  {The Journal of chemical physics}\ }\textbf {\bibinfo {volume} {150}},\
  \bibinfo {pages} {094509} (\bibinfo {year} {2019})}\BibitemShut {NoStop}%
\bibitem [{\citenamefont {Brotzakis}\ \emph {et~al.}(2019)\citenamefont
  {Brotzakis}, \citenamefont {Mendels},\ and\ \citenamefont
  {Parrinello}}]{brotzakis2019augmented}%
  \BibitemOpen
  \bibfield  {author} {\bibinfo {author} {\bibfnamefont {Z.~F.}\ \bibnamefont
  {Brotzakis}}, \bibinfo {author} {\bibfnamefont {D.}~\bibnamefont {Mendels}},\
  and\ \bibinfo {author} {\bibfnamefont {M.}~\bibnamefont {Parrinello}},\
  }\bibfield  {title} {\bibinfo {title} {Augmented harmonic linear discriminant
  analysis},\ }\href@noop {} {\bibfield  {journal} {\bibinfo  {journal} {arXiv
  preprint arXiv:1902.08854}\ } (\bibinfo {year} {2019})}\BibitemShut {NoStop}%
\bibitem [{\citenamefont {Piaggi}\ \emph {et~al.}(2017)\citenamefont {Piaggi},
  \citenamefont {Valsson},\ and\ \citenamefont
  {Parrinello}}]{piaggi2017enhancing}%
  \BibitemOpen
  \bibfield  {author} {\bibinfo {author} {\bibfnamefont {P.~M.}\ \bibnamefont
  {Piaggi}}, \bibinfo {author} {\bibfnamefont {O.}~\bibnamefont {Valsson}},\
  and\ \bibinfo {author} {\bibfnamefont {M.}~\bibnamefont {Parrinello}},\
  }\bibfield  {title} {\bibinfo {title} {Enhancing entropy and enthalpy
  fluctuations to drive crystallization in atomistic simulations},\ }\href@noop
  {} {\bibfield  {journal} {\bibinfo  {journal} {Phys. Rev. Lett.}\ }\textbf
  {\bibinfo {volume} {119}},\ \bibinfo {pages} {015701} (\bibinfo {year}
  {2017})}\BibitemShut {NoStop}%
\bibitem [{\citenamefont {Mendels}\ \emph
  {et~al.}(2018{\natexlab{c}})\citenamefont {Mendels}, \citenamefont {McCarty},
  \citenamefont {Piaggi},\ and\ \citenamefont {Parrinello}}]{Mendels2018}%
  \BibitemOpen
  \bibfield  {author} {\bibinfo {author} {\bibfnamefont {D.}~\bibnamefont
  {Mendels}}, \bibinfo {author} {\bibfnamefont {J.}~\bibnamefont {McCarty}},
  \bibinfo {author} {\bibfnamefont {P.~M.}\ \bibnamefont {Piaggi}},\ and\
  \bibinfo {author} {\bibfnamefont {M.}~\bibnamefont {Parrinello}},\ }\bibfield
   {title} {\bibinfo {title} {{Searching for Entropically Stabilized Phases:
  The Case of Silver Iodide}},\ }\href
  {https://doi.org/10.1021/acs.jpcc.7b11002} {\bibfield  {journal} {\bibinfo
  {journal} {J. Phys. Chem. C}\ }\textbf {\bibinfo {volume} {122}},\ \bibinfo
  {pages} {1786} (\bibinfo {year} {2018}{\natexlab{c}})}\BibitemShut {NoStop}%
\bibitem [{\citenamefont {Wodak}\ \emph {et~al.}(2019)\citenamefont {Wodak},
  \citenamefont {Paci}, \citenamefont {Dokholyan}, \citenamefont {Berezovsky},
  \citenamefont {Horovitz}, \citenamefont {Li}, \citenamefont {Hilser},
  \citenamefont {Bahar}, \citenamefont {Karanicolas}, \citenamefont {Stock}
  \emph {et~al.}}]{wodak2019allostery}%
  \BibitemOpen
  \bibfield  {author} {\bibinfo {author} {\bibfnamefont {S.~J.}\ \bibnamefont
  {Wodak}}, \bibinfo {author} {\bibfnamefont {E.}~\bibnamefont {Paci}},
  \bibinfo {author} {\bibfnamefont {N.~V.}\ \bibnamefont {Dokholyan}}, \bibinfo
  {author} {\bibfnamefont {I.~N.}\ \bibnamefont {Berezovsky}}, \bibinfo
  {author} {\bibfnamefont {A.}~\bibnamefont {Horovitz}}, \bibinfo {author}
  {\bibfnamefont {J.}~\bibnamefont {Li}}, \bibinfo {author} {\bibfnamefont
  {V.~J.}\ \bibnamefont {Hilser}}, \bibinfo {author} {\bibfnamefont
  {I.}~\bibnamefont {Bahar}}, \bibinfo {author} {\bibfnamefont
  {J.}~\bibnamefont {Karanicolas}}, \bibinfo {author} {\bibfnamefont
  {G.}~\bibnamefont {Stock}}, \emph {et~al.},\ }\bibfield  {title} {\bibinfo
  {title} {Allostery in its many disguises: from theory to applications},\
  }\href@noop {} {\bibfield  {journal} {\bibinfo  {journal} {Structure}\
  }\textbf {\bibinfo {volume} {27}},\ \bibinfo {pages} {566} (\bibinfo {year}
  {2019})}\BibitemShut {NoStop}%
\bibitem [{\citenamefont {Faure}\ \emph {et~al.}(2022)\citenamefont {Faure},
  \citenamefont {Domingo}, \citenamefont {Schmiedel}, \citenamefont
  {Hidalgo-Carcedo}, \citenamefont {Diss},\ and\ \citenamefont
  {Lehner}}]{faure2022mapping}%
  \BibitemOpen
  \bibfield  {author} {\bibinfo {author} {\bibfnamefont {A.~J.}\ \bibnamefont
  {Faure}}, \bibinfo {author} {\bibfnamefont {J.}~\bibnamefont {Domingo}},
  \bibinfo {author} {\bibfnamefont {J.~M.}\ \bibnamefont {Schmiedel}}, \bibinfo
  {author} {\bibfnamefont {C.}~\bibnamefont {Hidalgo-Carcedo}}, \bibinfo
  {author} {\bibfnamefont {G.}~\bibnamefont {Diss}},\ and\ \bibinfo {author}
  {\bibfnamefont {B.}~\bibnamefont {Lehner}},\ }\bibfield  {title} {\bibinfo
  {title} {Mapping the energetic and allosteric landscapes of protein binding
  domains},\ }\href@noop {} {\bibfield  {journal} {\bibinfo  {journal}
  {Nature}\ }\textbf {\bibinfo {volume} {604}},\ \bibinfo {pages} {175}
  (\bibinfo {year} {2022})}\BibitemShut {NoStop}%
\bibitem [{\citenamefont {Hunter}\ and\ \citenamefont
  {Anderson}(2009)}]{hunter2009cooperativity}%
  \BibitemOpen
  \bibfield  {author} {\bibinfo {author} {\bibfnamefont {C.~A.}\ \bibnamefont
  {Hunter}}\ and\ \bibinfo {author} {\bibfnamefont {H.~L.}\ \bibnamefont
  {Anderson}},\ }\bibfield  {title} {\bibinfo {title} {What is
  cooperativity?},\ }\href@noop {} {\bibfield  {journal} {\bibinfo  {journal}
  {Angewandte Chemie International Edition}\ }\textbf {\bibinfo {volume}
  {48}},\ \bibinfo {pages} {7488} (\bibinfo {year} {2009})}\BibitemShut
  {NoStop}%
\bibitem [{\citenamefont {Barducci}\ \emph {et~al.}(2008)\citenamefont
  {Barducci}, \citenamefont {Bussi},\ and\ \citenamefont
  {Parrinello}}]{Barducci2008}%
  \BibitemOpen
  \bibfield  {author} {\bibinfo {author} {\bibfnamefont {A.}~\bibnamefont
  {Barducci}}, \bibinfo {author} {\bibfnamefont {G.}~\bibnamefont {Bussi}},\
  and\ \bibinfo {author} {\bibfnamefont {M.}~\bibnamefont {Parrinello}},\
  }\bibfield  {title} {\bibinfo {title} {{Well-tempered metadynamics: A
  smoothly converging and tunable free-energy method}},\ }\href
  {https://doi.org/10.1103/PhysRevLett.100.020603} {\bibfield  {journal}
  {\bibinfo  {journal} {Phys. Rev. Lett.}\ }\textbf {\bibinfo {volume} {100}},\
  \bibinfo {pages} {020603} (\bibinfo {year} {2008})},\ \Eprint
  {https://arxiv.org/abs/0803.3861} {arXiv:0803.3861} \BibitemShut {NoStop}%
\bibitem [{\citenamefont {Tiwary}\ and\ \citenamefont
  {Parrinello}(2015)}]{tiwary2015time}%
  \BibitemOpen
  \bibfield  {author} {\bibinfo {author} {\bibfnamefont {P.}~\bibnamefont
  {Tiwary}}\ and\ \bibinfo {author} {\bibfnamefont {M.}~\bibnamefont
  {Parrinello}},\ }\bibfield  {title} {\bibinfo {title} {A time-independent
  free energy estimator for metadynamics},\ }\href@noop {} {\bibfield
  {journal} {\bibinfo  {journal} {The Journal of Physical Chemistry B}\
  }\textbf {\bibinfo {volume} {119}},\ \bibinfo {pages} {736} (\bibinfo {year}
  {2015})}\BibitemShut {NoStop}%
\bibitem [{\citenamefont {Parrinello}\ and\ \citenamefont
  {Rahman}(1982)}]{parrinello1982strain}%
  \BibitemOpen
  \bibfield  {author} {\bibinfo {author} {\bibfnamefont {M.}~\bibnamefont
  {Parrinello}}\ and\ \bibinfo {author} {\bibfnamefont {A.}~\bibnamefont
  {Rahman}},\ }\bibfield  {title} {\bibinfo {title} {Strain fluctuations and
  elastic constants},\ }\href@noop {} {\bibfield  {journal} {\bibinfo
  {journal} {The Journal of Chemical Physics}\ }\textbf {\bibinfo {volume}
  {76}},\ \bibinfo {pages} {2662} (\bibinfo {year} {1982})}\BibitemShut
  {NoStop}%
\bibitem [{\citenamefont {Parrinello}\ and\ \citenamefont
  {Rahman}(1981)}]{parrinello1981polymorphic}%
  \BibitemOpen
  \bibfield  {author} {\bibinfo {author} {\bibfnamefont {M.}~\bibnamefont
  {Parrinello}}\ and\ \bibinfo {author} {\bibfnamefont {A.}~\bibnamefont
  {Rahman}},\ }\bibfield  {title} {\bibinfo {title} {Polymorphic transitions in
  single crystals: A new molecular dynamics method},\ }\href@noop {} {\bibfield
   {journal} {\bibinfo  {journal} {Journal of Applied physics}\ }\textbf
  {\bibinfo {volume} {52}},\ \bibinfo {pages} {7182} (\bibinfo {year}
  {1981})}\BibitemShut {NoStop}%
\bibitem [{\citenamefont {Martyna}\ \emph {et~al.}(1994)\citenamefont
  {Martyna}, \citenamefont {Tobias},\ and\ \citenamefont
  {Klein}}]{martyna1994constant}%
  \BibitemOpen
  \bibfield  {author} {\bibinfo {author} {\bibfnamefont {G.~J.}\ \bibnamefont
  {Martyna}}, \bibinfo {author} {\bibfnamefont {D.~J.}\ \bibnamefont
  {Tobias}},\ and\ \bibinfo {author} {\bibfnamefont {M.~L.}\ \bibnamefont
  {Klein}},\ }\bibfield  {title} {\bibinfo {title} {Constant pressure molecular
  dynamics algorithms},\ }\href@noop {} {\bibfield  {journal} {\bibinfo
  {journal} {J. Chem. Phys.}\ }\textbf {\bibinfo {volume} {101}},\ \bibinfo
  {pages} {4177} (\bibinfo {year} {1994})}\BibitemShut {NoStop}%
\bibitem [{\citenamefont {Marto{\v{n}}{\'a}k}\ \emph
  {et~al.}(2003)\citenamefont {Marto{\v{n}}{\'a}k}, \citenamefont {Laio},\ and\
  \citenamefont {Parrinello}}]{martovnak2003predicting}%
  \BibitemOpen
  \bibfield  {author} {\bibinfo {author} {\bibfnamefont {R.}~\bibnamefont
  {Marto{\v{n}}{\'a}k}}, \bibinfo {author} {\bibfnamefont {A.}~\bibnamefont
  {Laio}},\ and\ \bibinfo {author} {\bibfnamefont {M.}~\bibnamefont
  {Parrinello}},\ }\bibfield  {title} {\bibinfo {title} {Predicting crystal
  structures: the parrinello-rahman method revisited},\ }\href@noop {}
  {\bibfield  {journal} {\bibinfo  {journal} {Physical review letters}\
  }\textbf {\bibinfo {volume} {90}},\ \bibinfo {pages} {075503} (\bibinfo
  {year} {2003})}\BibitemShut {NoStop}%
\bibitem [{\citenamefont {Liu}\ and\ \citenamefont
  {Nagel}(2010)}]{liu2010jamming}%
  \BibitemOpen
  \bibfield  {author} {\bibinfo {author} {\bibfnamefont {A.~J.}\ \bibnamefont
  {Liu}}\ and\ \bibinfo {author} {\bibfnamefont {S.~R.}\ \bibnamefont
  {Nagel}},\ }\bibfield  {title} {\bibinfo {title} {The jamming transition and
  the marginally jammed solid},\ }\href@noop {} {\bibfield  {journal} {\bibinfo
   {journal} {Annu. Rev. Condens. Matter Phys.}\ }\textbf {\bibinfo {volume}
  {1}},\ \bibinfo {pages} {347} (\bibinfo {year} {2010})}\BibitemShut {NoStop}%
\bibitem [{\citenamefont {Rocks}\ \emph {et~al.}(2019)\citenamefont {Rocks},
  \citenamefont {Ronellenfitsch}, \citenamefont {Liu}, \citenamefont {Nagel},\
  and\ \citenamefont {Katifori}}]{rocks2019limits}%
  \BibitemOpen
  \bibfield  {author} {\bibinfo {author} {\bibfnamefont {J.~W.}\ \bibnamefont
  {Rocks}}, \bibinfo {author} {\bibfnamefont {H.}~\bibnamefont
  {Ronellenfitsch}}, \bibinfo {author} {\bibfnamefont {A.~J.}\ \bibnamefont
  {Liu}}, \bibinfo {author} {\bibfnamefont {S.~R.}\ \bibnamefont {Nagel}},\
  and\ \bibinfo {author} {\bibfnamefont {E.}~\bibnamefont {Katifori}},\
  }\bibfield  {title} {\bibinfo {title} {Limits of multifunctionality in
  tunable networks},\ }\href@noop {} {\bibfield  {journal} {\bibinfo  {journal}
  {Proceedings of the National Academy of Sciences}\ }\textbf {\bibinfo
  {volume} {116}},\ \bibinfo {pages} {2506} (\bibinfo {year}
  {2019})}\BibitemShut {NoStop}%
\bibitem [{\citenamefont {Goodrich}\ \emph {et~al.}(2015)\citenamefont
  {Goodrich}, \citenamefont {Liu},\ and\ \citenamefont
  {Nagel}}]{goodrich2015principle}%
  \BibitemOpen
  \bibfield  {author} {\bibinfo {author} {\bibfnamefont {C.~P.}\ \bibnamefont
  {Goodrich}}, \bibinfo {author} {\bibfnamefont {A.~J.}\ \bibnamefont {Liu}},\
  and\ \bibinfo {author} {\bibfnamefont {S.~R.}\ \bibnamefont {Nagel}},\
  }\bibfield  {title} {\bibinfo {title} {The principle of independent
  bond-level response: Tuning by pruning to exploit disorder for global
  behavior},\ }\href@noop {} {\bibfield  {journal} {\bibinfo  {journal}
  {Physical review letters}\ }\textbf {\bibinfo {volume} {114}},\ \bibinfo
  {pages} {225501} (\bibinfo {year} {2015})}\BibitemShut {NoStop}%
\bibitem [{\citenamefont {Plimpton}(1995)}]{PLIMPTON1995}%
  \BibitemOpen
  \bibfield  {author} {\bibinfo {author} {\bibfnamefont {S.}~\bibnamefont
  {Plimpton}},\ }\bibfield  {title} {\bibinfo {title} {Fast parallel algorithms
  for short-range molecular dynamics},\ }\href
  {https://doi.org/http://dx.doi.org/10.1006/jcph.1995.1039} {\bibfield
  {journal} {\bibinfo  {journal} {J. Comput. Phys.}\ }\textbf {\bibinfo
  {volume} {117}},\ \bibinfo {pages} {1 } (\bibinfo {year} {1995})}\BibitemShut
  {NoStop}%
\bibitem [{\citenamefont {Tribello}\ \emph {et~al.}(2014)\citenamefont
  {Tribello}, \citenamefont {Bonomi}, \citenamefont {Branduardi}, \citenamefont
  {Camilloni},\ and\ \citenamefont {Bussi}}]{tribello_2014}%
  \BibitemOpen
  \bibfield  {author} {\bibinfo {author} {\bibfnamefont {G.~A.}\ \bibnamefont
  {Tribello}}, \bibinfo {author} {\bibfnamefont {M.}~\bibnamefont {Bonomi}},
  \bibinfo {author} {\bibfnamefont {D.}~\bibnamefont {Branduardi}}, \bibinfo
  {author} {\bibfnamefont {C.}~\bibnamefont {Camilloni}},\ and\ \bibinfo
  {author} {\bibfnamefont {G.}~\bibnamefont {Bussi}},\ }\bibfield  {title}
  {\bibinfo {title} {{PLUMED 2: New feathers for an old bird}},\ }\href@noop {}
  {\bibfield  {journal} {\bibinfo  {journal} {Comput. Phys. Commun.}\ }\textbf
  {\bibinfo {volume} {185}},\ \bibinfo {pages} {604} (\bibinfo {year}
  {2014})}\BibitemShut {NoStop}%
\bibitem [{\citenamefont {Schneider}\ and\ \citenamefont
  {Stoll}(1978)}]{schneider1978molecular}%
  \BibitemOpen
  \bibfield  {author} {\bibinfo {author} {\bibfnamefont {T.}~\bibnamefont
  {Schneider}}\ and\ \bibinfo {author} {\bibfnamefont {E.}~\bibnamefont
  {Stoll}},\ }\bibfield  {title} {\bibinfo {title} {Molecular-dynamics study of
  a three-dimensional one-component model for distortive phase transitions},\
  }\href@noop {} {\bibfield  {journal} {\bibinfo  {journal} {Physical Review
  B}\ }\textbf {\bibinfo {volume} {17}},\ \bibinfo {pages} {1302} (\bibinfo
  {year} {1978})}\BibitemShut {NoStop}%
\bibitem [{\citenamefont {Parrinello}\ and\ \citenamefont
  {Rahman}(1980)}]{parrinello1980crystal}%
  \BibitemOpen
  \bibfield  {author} {\bibinfo {author} {\bibfnamefont {M.}~\bibnamefont
  {Parrinello}}\ and\ \bibinfo {author} {\bibfnamefont {A.}~\bibnamefont
  {Rahman}},\ }\bibfield  {title} {\bibinfo {title} {Crystal structure and pair
  potentials: A molecular-dynamics study},\ }\href@noop {} {\bibfield
  {journal} {\bibinfo  {journal} {Phys. Rev. Lett.}\ }\textbf {\bibinfo
  {volume} {45}},\ \bibinfo {pages} {1196} (\bibinfo {year}
  {1980})}\BibitemShut {NoStop}%
\end{thebibliography}%


\newpage
\section*{\textsc{Supporting information} }


\renewcommand{\thepage}{S\arabic{page}} 
\renewcommand{\thefigure}{S\arabic{figure}}
\renewcommand{\thetable}{S\arabic{table}}
\renewcommand{\theequation}{S\arabic{equation}}







\subsection*{\noindent Rotation of the 2d plane spanned by the top HLDA eigenvectors}

\begin{figure}[!htbp]
	\centering
    \includegraphics[trim=0 90 80 50,clip, width=0.96\columnwidth]{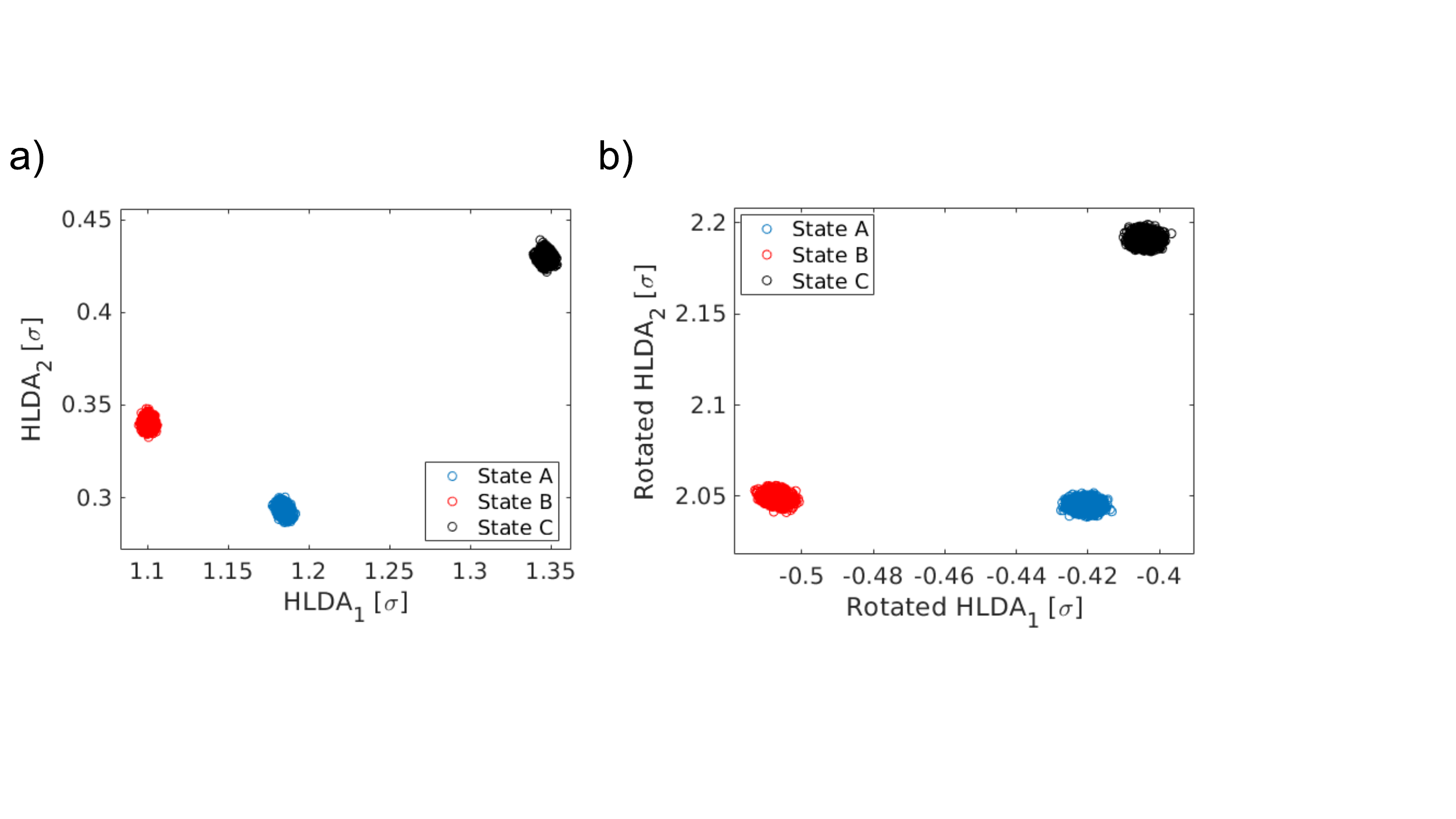}   
    \caption{ a) Projection of the data sampled in the training simulations of the three considered states (A being the inactivated state, B the activated state, and C the inhibiting state) on the plane spanned by the top two HLDA eigenvectors. b) The HLDA eigenvector plane after rotation. It can be seen that states A and B are now predominantly separated with respect to the rotated HLDA\textsubscript{2}, while states A and C are predominantly separated with respect to the rotated HLDA\textsubscript{1}. 
    }
    \label{fig:S1}
\end{figure}



\subsection*{\noindent Example of time dependence behavior in a WTMD simulation in which $d_s$ is the biased CV}

\begin{figure}[!htbp]
	\centering
    \includegraphics[trim=0 0 275 0,clip, width=0.66\columnwidth]{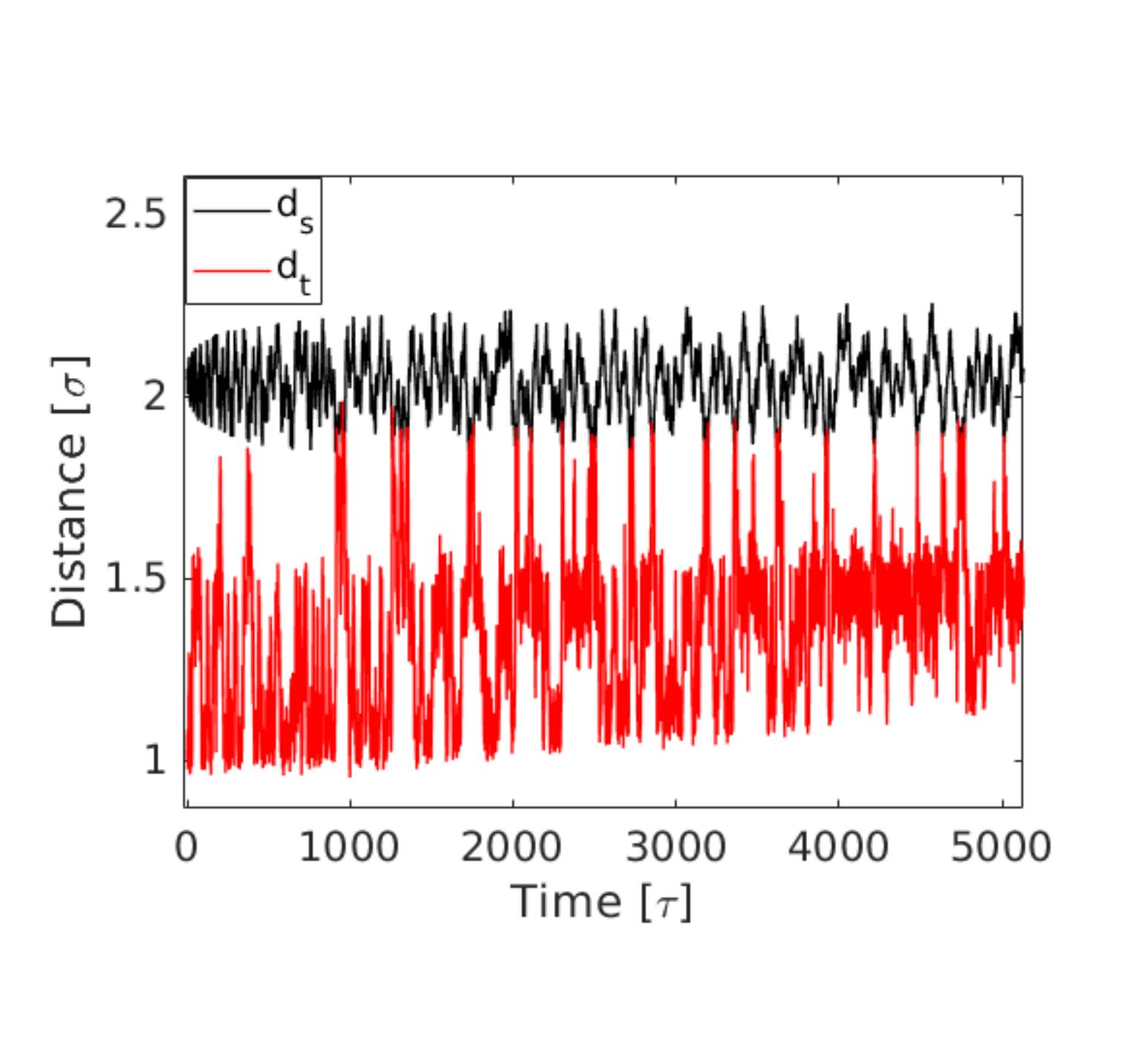}   
    \caption{Time dependence excerpt from a WTMD simulation run at $T=8.6 \cdot 10^{-6}$ ($k_B=1$) in which $d_s$, the distance between the source nodes is the biased CV. It can be seen that upon passing a threshold value, the allosteric response of the target beads (measured by the distance between them, $d_t$) is triggered.
    }
    \label{fig:S2}
\end{figure}

\newpage
\subsection*{\noindent Time dependence behavior in WTMD simulations for a pristine and modified network }

\begin{figure}[!htbp]
	\centering
    \includegraphics[trim=0 30 120 0,clip, width=0.96\columnwidth]{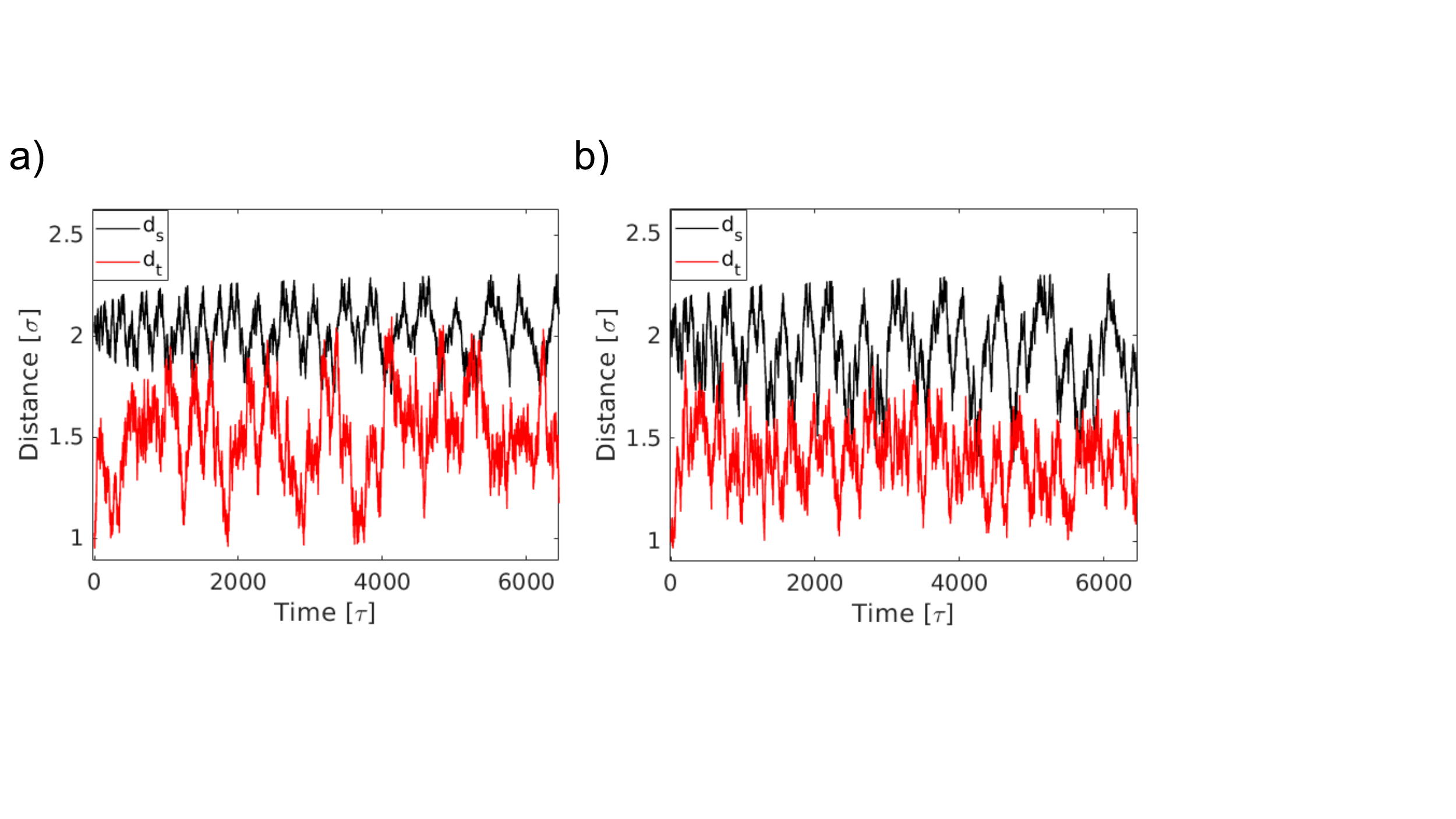}   
    \caption{a) Time dependence excerpt from a WTMD simulation of the pristine network run at $T=4.3 \cdot 10^{-5}$ in which $d_s$, the distance between the source nodes is the biased CV. It can be seen that upon passing a threshold value, the allosteric response of the target beads (measured by the distance between them, $d_t$) is triggered. b) Time dependence excerpt from a WTMD simulation run at $T=4.3 \cdot 10^{-5}$ of a modified network for which the bonds ranked 12-27 in the HLDA\textsubscript{1} weight hierarchy were weakened. As can be seen the activated state is inhibited in the modified network whereby for values $d_s<1.7$ the target beads do not access the activated region of $d_t>1.7$ 
    }
    \label{fig:DeltaF_protein_for_SI_angles}
\end{figure}


\subsection*{\noindent Equation for the calculation of the free energy difference $\Delta F$ between any two states A and B}

\begin{equation}
\label{DeltaF}
\Delta F_{AB}=-\frac{1}{\beta} \log{\frac{\int_A d\mathbf{s} e^{-\beta F(\mathbf{s})}}{\int_B d\mathbf{s} e^{-\beta F(\mathbf{s})}}}
\end{equation}



\end{document}